\documentstyle{npbproc}

\def\spose#1{\hbox to 0pt{#1\hss}}

\font\amsx=msam10
\def\bbox{\hbox{\amsx\char'004}}
\def\gsim{\hbox{\amsx\char'046}}
\def\btri{\hbox{\amsx\char'116}}
\font\amsy=msbm10
\def\II{\hbox{\amsy\char'111}}
\def\ptl{\partial}
\def\curl{\mathop{\rm curl}\nolimits}
\def\tr{\mathop{\rm tr}\nolimits}
\def\ov#1{\overline{#1}}
\def\wt#1{\widetilde{#1}}
\def\eqd{\mathop{\ \spose{\lower 5pt\hbox{\hskip 1pt{\sixrm def}}}=}\ }
\def\hangref#1{\noindent\hangindent 20pt #1\par}
\def\blbox{\raise 1pt\hbox{\hskip 1pt\vrule width 3.5pt height 3.5pt
  \hskip 1pt}}
\def\disp#1{\relax\begin{displaymath} #1\end{displaymath}}
\def\equa#1{\relax\begin{equation} #1\end{equation}}
\def\epsi{\varepsilon}

\begin{document}
\title{Multigrid Methods in Lattice Field Computations}
\author{Achi Brandt}
\address{Department of Applied Mathematics and Computer Science\\
The Weizmann Institute of Science\\
Rehovot, Israel 76100}
\date{}
\runtitle{Multigrid Methods in Lattice Field Computations}
\runauthor{A.\ Brandt}
\volume{XXX}
\firstpage{1}
\lastpage{44}
\begin{abstract}
The multigrid methodology is reviewed. By integrating numerical processes
at all scales of a problem, it seeks to perform various computational
tasks at a cost that rises as slowly as possible as a function of $n$,
the number of degrees of freedom in the problem. Current and potential
benefits for lattice field computations are outlined. They include:
$O(n)$ solution of Dirac equations; just $O(1)$ operations in updating
the solution (upon any local change of data, including the gauge field);
similar efficiency in gauge fixing and updating; $O(1)$ operations in
updating the inverse matrix and in calculating the change in the
logarithm of its determinant; $O(n)$ operations per producing each
independent configuration in statistical simulations (eliminating CSD),
and, more important, effectively just $O(1)$ operations per each
independent measurement (eliminating the volume factor as well). These
potential capabilities have been demonstrated on simple model problems.
Extensions to real life are explored.
\end{abstract}
\maketitle
\section{Elementary Acquaintance with Multigrid}
\subsection{Particle minimization problem}
To introduce some of the basic concepts of multi-scale computations,
consider the simple example where one wishes to calculate the effect of
an external field on the stationary state of a piece of solid made of $n$
classical atoms. Denote by $r_i=(r_{i1},r_{i2},r_{i3})$ the coordinates
of the $i$-th atom, by $r=(r_1,r_2,\ldots,r_n)$ the vector of all atom
positions (the configuration) and by
\def\theequation{1.1}
\equa{E^0(r)=\sum V_{ij}(\vert r_i-r_j\vert)}
the energy of their mutual interactions. Let $r^0=(r_1^0,r_2^0,\ldots,
r_n^0)$ be the given steady state in the absence of an external field,
i.e.,
\disp{E^0(r^0)=\min_r E^0(r),}\noindent
entailing $\ptl E^0(r^0)/\ptl r_{i\mu}=0$ $(i=1,2,\ldots,n;\ \mu=1,2,3)$.
One wishes to calculate the state $r^{\ast}$ obtained when external
forces $f=(f_1,f_2,\ldots,f_n)$ are added, i.e.,
\def\theequation{1.2}
\equa{E(r^{\ast})=\min_r E(r)}
where, for example,
\def\theequation{1.3}
\equa{E(r)=E^0(r)-\sum_i f_ir_i.}
\par\vskip -\parskip
The computational problem of fast {\it evaluation\/}, for any given $r$,
of $E(r)$, or of the {\it residual forces\/}
\disp{\nabla E(r)=\biggl({\ptl E\over\ptl r_{i\mu}}(r);\ i=1,2,\ldots,n;\
\mu=1,2,3\biggr)}\noindent
in case {\it far\/} interactions are significant (e.g., electrostatic
forces) is mentioned in Sec.~6. Here we confine our attention to the
problem of finding $r^{\ast}$.\par
\subsection{Relaxation}
A general approach for calculating $r^{\ast}$ is the {\it
particle-by-particle minimization\/} or {\it relaxation\/}. At each step
of this process, the position of only one particle, $r_i$ say, is
changed, keeping the position of all others fixed. The new value of $r_i$
is chosen so as to reduce $E(r)$ as much as possible. Repeating this step
for all particles $(i=1,2,\ldots,n)$ is called a {\it relaxation
sweep\/}. By performing a sufficiently long sequence of relaxation
sweeps, one hopes to be able to get as close to $r^{\ast}$ as one wishes.
\par
One main difficulty with the relaxation process is its extremely slow
convergence. The reason is that, when all other particles are held fixed,
$r_i$ can change only slightly, only a fraction of the typical distance
between neighboring atoms, before the energy $E(r)$ starts to rise (very
sharply). Hence, very many relaxation sweeps will be needed to
obtain a new configuration $r$ (a new shape of the solid) macroscopically
different from $r^0$. (Another possible difficulty --- the danger of
converging to the wrong solution --- is related to global optimization;
cf.\ Sec.~6).\par
To be sure, if the external forces on neighboring particles are very
different from each other, the first few relaxation sweeps may exhibit
fast local adjustments of the configuration, rapidly yielding
configurations with possibly much smaller residual forces $\nabla E(r)$.
But the advance thereafter toward large-scale changes will be painfully
slow, eventually exhibiting also very slow further reduction of the
residual forces. The slowness clearly increases with the size $n$: the
more atoms in the system, the more relaxation {\it sweeps\/} that are
needed to achieve reasonable convergence.\par
\subsection{Multiscale relaxation}
Moving only one particle at a time being so inefficient, ways are
evidently needed to perform {\it collective\/} motions of atoms.\par
A {\it collective displacement on scale\/} $h$, say, with {\it center\/}
$x_k=(x_{k1},x_{k2},x_{k3})$, and {\it amplitude\/} $u_k=(u_{k1},u_{k2},
u_{k3})$ and {\it shape function\/} $w_k(\zeta)$ can be defined by the
replacement
\disp{r_i\leftarrow r_i+\delta r_i,\qquad (1\le i\le n)}\noindent
where
\def\theequation{1.4}
\equa{\delta r_i=w_k\biggl({r_i-x_k\over h}\biggr)u_k.}
The shape function $w_k$ is chosen so that $w_k(0)=1$, while $w_k(\zeta)=
0$ for all $\vert\zeta\vert\ge C$, where $\zeta=(\zeta_1,\zeta_2,\zeta_3
)$, $\vert\zeta\vert=\max(\vert\zeta_1\vert,\vert\zeta_2\vert,\vert
\zeta_3\vert)$ and $C$ is a small integer (often $C=1$). Hence $\delta
r_i=0$ for $\vert r_i-x_k\vert\ge Ch$; i.e., only atoms at distance
$O(h)$ from the center are actually moved. The shape function can often
be chosen independently of $k$; a typical shape is the ``pyramid" $w_k(
\zeta)=1-\vert\zeta\vert$. The displacement described by it affects only
atoms occupying a $2h\times 2h\times 2h$ cube; it will leave all of them
within that cube as long as $\vert u_k\vert=\max(\vert u_{k1}\vert,\vert
u_{k2}\vert,\vert u_{k3}\vert\le h$.\par
{\it A scale\/ $h$ relaxation step\/} is performed at a point $x_k$ by
choosing $u_k$ so as to reduce $E(r+\delta r)$ as far as possible (or as
far as convenient and practical to inexpensively calculate. Since this is
only one step in an iterative process, it would often be a major waste of
effort to calculate that $u_k$ which actually minimizes $E(r+\delta r)$.)
A {\it relaxation sweep on scale\/} $h$ is a sequence of such steps, with
$x_k$ scanning the gridpoints $x_1,x_2,\ldots$ of a grid (lattice) with
meshsize $h$ placed over the domain occupied by the atoms.\par
What scale $h$ should be chosen for the movements? For movements on a
small scale $h$, comparable to the inter-atomic distances, a slow-down
similar to that of the particle-by-particle minimization will take place.
Indeed, to reduce the energy, $\vert u_k\vert$ must be smaller (usually
substantially smaller) than $h$. Large values of $h$, approaching the
linear size of the domain, will allow large scale movements, but will
fail to perform efficiently intermediate-scale movements. Such
intermediate scale movements are usually necessary since there is no
cheap way to choose shape functions that will exactly fit the required
large scale movement (which of course depends on the external field $f$).
Relaxation sweeps are thus needed at scales approximating {\it all\/}
scales of the problem.\par
A {\it multiscale relaxation cycle\/} is a process which includes
particle-by-particle relaxation sweeps plus relaxation sweeps on scales
$a,2a,4a,\ldots,2^la$, where $a$ is comparable to the inter-atomic
distance and $2^la$ is comparable to the radius of the domain. The
$1\colon 2$ meshsize ratio of such a cycle is tight enough to allow
efficient generation of all smooth movements of the atoms, even with a
fixed and simple (e.g., pyramidal) shape function. Such a cycle will not
slow down. If every cycle involves a couple of relaxation sweeps at every
level, the error (or the residual forces) will typically drop by an order
of magnitude per cycle!\par
{\it A comment on names\/}: What we call here ``multiscale relaxation" is
equivalent to a process which is called ``multigrid" by some recent
authors (see Sec.\ 5.1), but which has actually been named ``unigrid" in
the traditional multigrid literature \cite{UG1}, \cite{UG2}, \cite{UG3},
since all its moves are still being explicitly performed in terms of the
finest level (here: the level of moving single atoms). The term
``multigrid" traditionally implies some additional important ideas, to be
discussed next.\par
\subsection{Displacement fields}
Instead of performing one displacement at a time on the ensemble of
particles, it will be more efficient (and important in other ways that
will be explained in Sec.\ 5.1 below) to regard the set of displacement
amplitudes $u_1^h,u_2^h,\ldots$ (at respective centers $x_1^h,x_2^h,
\ldots$ forming a grid with meshsize $h$) as one field, to be jointly
calculated. Instead of (1.4), the field $\delta r$ of particle moves will
then be given by
\def\theequation{1.5}
\equa{\delta r_i=\sum_k w_k^h\biggl({r_i-x_k^h\over h}\biggr)u_k^h.}
Note the superscripts $h$ added here to emphasize that $x_k^h$, $u_k^k$
and $w_k^h$ are in principle different on different grids $h$, although
the shapes $w_k^h$ will often be independent of both $k$ and $h$. The
choice of $w_k^h$ will normally be such that
\def\theequation{1.6}
\equa{\sum_k w_k^h\biggl({r_i-x_k^h\over h}\biggr)=1\ \hbox{for any}\ r_i
,}
so that (1.5) is simply an {\it interpolation\/} of the {\it displacement
field\/} $u^h=(u_1^h,u_2^h,\ldots)$ from the gridpoints to the (old)
particle locations.\par
For example with the choice
\def\theequation{1.7}
\equa{w_k^h(\zeta)=(1-\vert\zeta_1\vert)(1-\vert\zeta_2\vert)(1-\vert
\zeta_3\vert),}
relation (1.5) expresses the usual {\it tri-linear\/} interpolation: a
linear interpolation in each of the three coordinate directions,
performed successively in any order. This is actually a {\it second\/}
order interpolation; we generally say that the interpolation is {\it of
order\/} $p$ if, for any sufficiently smooth function $U(x)$,
\def\theequation{1.8}
\equa{U(x)=\sum_k w_k^h\biggl({x-x_k^h\over h}\biggr)U(x_k^h)+O(h^p).}
It is easy to see that (1.6) is necessary and sufficient for the
interpolation order to be at least 1.\par
Note that the pyramidal shape $w_k^h(\zeta)=1-\vert\zeta\vert$ will {\it
not\/} yield an interpolation, and is therefore inadequate here. For
example, for a {\it constant\/} displacement field ($u_k^h$ independent
of $k$) it will give sharply {\it variable\/} atomic moves.\par
For any fixed particle configuration $r$, the energy $E(r+\delta r)$ is,
by (1.5), a functional of the displacement field $u^h$; we denote this
functional $E^h$:
\disp{E^h(u^h)=E(r+\delta r).}\noindent
In principle one needs to perform the interpolation (1.5) in order to
evaluate $E^h(u^h)$. Usually, however, simpler and more explicit
evaluation, or approximate evaluation, procedures can be constructed.
This is especially possible if $h$ is small, comparable to the distances
between neighboring atoms ($h=a$, the finest scale in Sec.\ 1.3), because
on such a scale one can assume $u^h$ to be smooth, since non-smooth
motions have already been efficiently performed at the particle level (by
the particle-by-particle relaxation sweeps; these motions are the ``fast
local adjustments" mentioned in Sec.\ 1.2). For such smooth $u^h$ the
``strains", i.e., the differences $u_k^h-u_l^h$ between the displacements
at neighboring sites $x_k^h$ and $x_l^h$, are small. Since the change
$\delta r_i-\delta r_j$ in the relative position of a pair of neighboring
atoms $i$ and $j$ can, by (1.5) and (1.6), be written as a linear
combination of these small strains, the change in their potential
\disp{\delta V_{ij}=V_{ij}\bigl(\vert r_i+\delta r_i-(r_j+\delta r_j)
\vert\bigr)-V_{ij}(\vert r_i-r_j\vert)}\noindent
can be expanded in a Taylor series in terms of the small strains:
\def\theequation{1.9}
\equa{\delta V_{ij}\approx\sum^M_{m=1}\sum_{k,l} A_{ijklm}(u_k^h-u_l^h
)^m}
where the summation $\sum_{k,l}$ is over sites $x_k$ and $x_l$ in the
vicinity of $r_i$ and $r_j$. The degree $M$ of the expansion depends on
the desired accuracy. Since the displacement will be part of a
self-correcting iterative process, the minimal order $M=2$ will usually
suffice; higher orders would often be just a waste of effort. The
coefficients $A_{ijklm}$ depend of course on the base configuration $r$.
\par
{}From (1.3), (1.1) and (1.9), or by some other approximation, one obtains
$E(r+\delta r)$ as an {\it explicit\/} functional, $E^h(u^h)$, of the
displacement field. Hence, before ever returning to the particles
themselves, one can (approximately) calculate the displacement field
which would reduce the energy as far as possible, i.e., the values
$u^{h\ast}$ for which
\def\theequation{1.10}
\equa{E^h(u^{h\ast})=\min_{u^h} E^h(u^h).}
\par\vskip -\parskip
The fast (approximate) solution of (1.10) (i.e., finding $u^{h\ast}$) is
a {\it lattice\/} problem, to be discussed in the next section. As we
will see, the process will include, even though indirectly, displacements
on all coarser scales $2h,4h,\ldots$~.\par
Having calculated a field $u^h$ which approximates $u^{h\ast}$, one can
then return to the particles and displace all of them simultaneously,
using (1.5). This collective motion would introduce accurately the main
large-scale smooth change needed to approach the ground state $r^{\ast}$.
The remaining error would usually be non-smooth, hence quickly removable
by a couple of additional particle-by-particle relaxation sweeps. When
such sweeps start to indicate slow convergence (e.g., slow reduction of
residual forces), the small remaining error is again smooth, hence it can
again be substantially reduced by a {\it new\/} displacement field $u^h$,
(approximately) minimizing a {\it new\/} energy $E^h(u^h)$, similar to
the above, but constructed with respect to the new (the latest)
configuration $r$. And so on. Each such cycle, including a number of
particle-by-particle relaxation sweeps plus forming (1.9), solving Eq.\
(1.10) and moving according to (1.5), will typically reduce the error by
an order of magnitude.\par
\subsection{Lattice minimization problem}
Instead of the particle problem (1.2)--(1.3), consider now the analogous
lattice problem of finding the configuration $u^{\alpha\ast}$ which
minimizes the energy
\def\theequation{1.11}
\equa{E^{\alpha}(u^{\alpha})=\sum_{i,j}V_{ij}(u_i^{\alpha}-u_j^{\alpha})-
\sum_i f_i^{\alpha}u_i^{\alpha},}
where each $u_i^{\alpha}$ is a (scalar or vector) variable located at a
gridpoint $x_i^{\alpha}$ of a $d$-dimensional grid with meshsize
$\alpha$. The lattice energy $E^{\alpha}(u^{\alpha})$ can arise either as
the discretization of a continuum field energy, or as the particle
displacement energy described above ((1.10), with $h=\alpha$). (For more
comments on the relation between particle and continuum problems see
Sec.~6 below.)\par
Methods for minimizing (1.11) are fully analogous to those described
above for the particle problem. The basic step is to change one variable
$u_i^{\alpha}$ so as to reduce $E^{\alpha}(u^{\alpha})$, all other
variables being held fixed. Repeating such a step at every gridpoint
$x_i^{\alpha}$ is called a {\it point-by-point relaxation sweep\/}. Such
a sweep may initially rapidly reduce the {\it residuals\/} $\ptl
E^{\alpha}/\ptl u_i^{\alpha}$, but as soon as the error $u^{\alpha\ast}-
u^{\alpha}$ becomes smooth, the convergence will become very slow. In
fact, {\it no\/} local processing can efficiently reduce smooth errors,
because their size depends on more global information (e.g., on the
residuals over a domain substantially larger than the domain over which
the residuals have one dominant sign, or one dominant direction). Thus,
steps of more global nature are required.\par
{\it Relaxation on scale\/} $h$ can be devised here in the same manner as
above, based on moves
\def\theequation{1.12}
\equa{\delta u_i^{\alpha}=w_k\biggl({x_i^{\alpha}-x_k\over h}\biggr)u_k,}
with the same shape function $w_k(\zeta)$ as used in (1.4). {\it A
relaxation sweep on scale\/} $h$ is repeating (1.12) with $x_k$ scanning
all the gridpoints of a grid with meshsize $h$. A {\it multiscale
relaxation cycle\/} is a process that typically includes a couple of
relaxation sweeps on each of the scales $\alpha,2\alpha,\ldots,2^l
\alpha$, where a relaxation on the scale $\alpha$ is just the
point-by-point relaxation mentioned before, and $2^l\alpha$ is a meshsize
comparable to the linear size of the entire domain. (Usually, taking
every other hyperplane of a grid $h$ will give the hyperplanes of the
grid $2h$. In case of rectangular or periodic domains, the finest
meshsize $\alpha$ is customarily chosen so that the original grid has a
multiple of $2^{\alpha}$ intervals in each direction, to maintain
simplicity at the coarser levels.) Each such cycle will typically reduce
the error by an order of magnitude.\par
\subsection{Multigrid cycles}
Instead of performing all the moves directly in terms of the finest grid,
as in (1.12), it will usually be more efficient and advantageous (see
Sec.\ 5.1) to consider the moves $u^h=(u_1^h,u_2^h,\ldots,u_{n^h}^h)$ for
grid $h$ (at its gridpoints $x_1^h,x_2^h,\ldots,x_{n^h}^h$, respectively)
as a {\it field\/} which {\it jointly\/} describes displacements for the
next finer field, $u^{h/2}$, via the relation
\def\theequation{1.13}
\equa{\delta u_i^{h/2}=\sum_k w_k^h\biggl({x_i^{h/2}-x_k^h\over h}\biggr)
u_k^h,}
where the shape functions satisfy (1.6) (or even (1.8), for some chosen
order $p$), so that (1.13) is in fact a proper interpolation (of order
$p$). We will denote this interpolation from grid $h$ to grid $h/2$ by
\def\theequation{1.14}
\equa{\delta u^{h/2}=I_h^{h/2}u^h.}
Each field $u^h$ will be governed by its own energy functional $E^h(u^h
)$, constructed from (1.14) and the relation
\def\theequation{1.15}
\equa{E^h(u^h)\approx E^{h/2}(u^{h/2}+\delta u^{h/2}),}
approximations (wherever needed to obtain a simple {\it explicit\/}
dependence of $E^h$ on $u^h$) being derived in the same manner as in
Sec.\ 1.4 above. Note that $E^h$ is constructed with respect to a given,
fixed configuration $u^{h/2}$; its coefficients are re-derived each time
the algorithm switches from moves on grid $h/2$ to moves on grid $h$.\par
Indeed, this recursive structure of fields and energies is operated by a
recursive algorithm. A {\it multigrid cycle for grid\/} $h/2$ is
recursively defined as consisting of the following 5 steps:\par
\hangref{(i) {\it Pre-relaxation\/}. Perform $\nu_1$ relaxation sweeps on
grid $h/2$.}\par
\hangref{(ii) {\it Coarsening\/}. With respect to the current
configuration $u^{h/2}$, construct the energy functional $E^h$, using
(1.15).}\par
\hangref{(iii) {\it Recursion\/}. If $h$ is the coarsest grid, minimize
$E^h(u^h)$ directly: the number of variables is so small that this should
be easy to do. Else, perform $\gamma$ multigrid cycles for grid $h$,
starting with the trivial initial approximation $u^h=0$.}\par
\hangref{(iv) {\it Uncoarsening\/}. Replace $u^{h/2}$ by $u^{h/2}+\delta
u^{h/2}$, using (1.13) with the final configuration $u^h$ obtained by
Step (iii).}\par
\hangref{(v) {\it Post-relaxation\/}. Perform $\nu_2$ additional
relaxation sweeps on grid $h/2$.}\par
The parameter $\gamma$ is called the {\it cycle index\/}; in minimization
problems it is usually 1 or 2. (Much larger indices may be used in
Monte-Carlo processes; see Sec.\ 5.4 below). When $\gamma=1$ the cycle is
called a $V$ cycle --- or $V(\nu_1,\nu_2)$, showing the number of pre-
and post-relaxation sweeps. For $\gamma=2$ the cycle is called a $W$
cycle, or $W(\nu_1,\nu_2)$. Fig.~1 displays the graphic origin of these
names.\par
\begin{figure*}[htb]
\vspace*{10truecm}
\caption{{\it Multigrid cycles\/}. A full circle $\bullet$ stands for
$\nu_1$ relaxation sweeps (pre-relaxation) on the level shown. Empty
circle $\circ$ indicates $\nu_2$ sweep (post-relaxation). A downward
arrow $\searrow$ means coarsening (calculation of the Hamiltonian $E^h$
from $E^{h/2}$). An upward arrow $\nearrow$ stands for uncoarsening
(displacing the field $u^{h/2}$ by the interpolated $u^h$).}
\end{figure*}%
{\it A multigrid solver\/} can consist of a sequence of multigrid cycles
for the finest grid $\alpha$. Each multigrid cycle, with $\nu_1+\nu_2=2$
or 3, will typically reduce the error by an order of magnitude. Since the
work on each level $h$ is only a fraction $(\gamma 2^{-d})$ of the work
on the next finer level $(h/2)$, most of the work in a cycle are the
$\nu_1+\nu_2$ point-by-point relaxation sweeps performed at the finest
grid $\alpha$. (An improved multigrid solver --- the FMG algorithm --- is
described in Sec.\ 3.1 below).\par
We have seen here an example of a fast multigrid solver. As we will see
below, multigrid-like structures and algorithms can serve many other
computational tasks.\par
\section{Introduction}
Multiscale (or ``multilevel" or ``multigrid") methods are techniques for
the fast execution of various many-variable computational
tasks defined in the physical space (or space-time, or any other similar
space). Such tasks include the solution of many-unknown equations (e.g.,
discretized partial differential and integral equations), the
minimization or statistical or dynamical simulations of many-particle or
large-lattice systems, the calculation of determinants, the derivation of
macroscopic equations from microscopic laws, and a variety of other
tasks. The multiscale algorithm includes local processing (relaxing an
equation, or locally reducing the energy, or simulating a local
statistical relation) at each scale of the problem together with
inter-scale interactions: typically, the evolving solution on each scale
recursively dictates the {\it equations\/} (or the Hamiltonian) on
coarser scales and modifies the {\it solution\/} (or configuration) on
finer scales. In this way large-scale changes are effectively performed
on coarse grids based on information gathered from finer grids.\par
As a result of such multilevel interactions, the fine scales of the
problem can be employed very sparingly, and sometimes only at special
and/or representative small regions. Moreover, the inter-scale
interactions can eliminate all kinds of troubles, such as: slow
convergence (in minimization processes, PDE solvers, etc.); critical
slowing down (in statistical physics); ill-posedness (e.g., of inverse
problems); large-scale attraction basins (in global optimization);
conflicts between small-scale and large-scale representations (e.g., in
wave problems); numerousness of interactions (in many body problems or
integral equations); the need to produce many fine-level solutions (e.g.,
in optimal control) or very many fine-level independent samples (in
statistical physics); etc.\par
The first multiscale algorithm was probably Southwell's two-level ``group
relaxation" for solving elliptic partial differential equations
\cite{A40}, first extended to more levels by Fedorenko \cite{A23}. These
and other early algorithms did not attract users because they lacked
understanding of the very local role to which the fine-scale processing
should be limited and of the real efficiency that can be attained by
multigrid and how to obtain it (e.g., at what meshsize ratio). The first
multigrid solvers exhibiting the generality and the typical modern
efficiency (e.g., four orders of magnitude faster than Fedorenko's
estimates) were developed in the early 1970's \cite{A3}, \cite{A9},
leading to extensive activity in this field. Much of this activity is
reported in the multigrid books \cite{A26}, \cite{A35}, \cite{A10},
\cite{A38}, \cite{A25}, \cite{A32}, \cite{A27}, \cite{A22}, \cite{A33},
\cite{A31}, \cite{A34}, \cite{A28}, \cite{A29} and references therein.
Recent developments, including in particular the development of
multiscale methods outside the field of partial differential equations,
are reviewed in \cite{A5}, \cite{A7}.\par
This article will review some of the basic {\it conceptual\/}
developments, with special emphasis on those relevant to lattice field
calculations. Sec.~3 will deal with the most developed area, that of
multigrid solvers for discretized partial differential equations; the
solvers for the special case of Dirac equation are discussed in more
detail in Sec.~4, including new approaches appearing here for the first
time (Sec.\ 4.7).\par
In most problem environments, the multi-level approach can give much more
than just a fast solver. For example, it yields very fast procedures for
updating the solution upon local changes in the data: see Sec.\ 4.8. For
LGT problems it can also provide very economic and rapidly updatable data
structures for storing inverse matrix information which allows immediate
updating of various required quantities, such as determinant values
needed for the fermionic interaction: see Sec.\ 4.9.\par
In {\it statistical\/} field computations, in addition to such fast
calculation with the fermionic matrix, multiscale processes can
potentially contribute in the following three ways (first outlined in
\cite{A5}). Firstly, in the same way that they accelerate minimization
processes (see Sec.~1 above), they can eliminate the ``critical slowing
down" in statistical simulations: see Secs.\ 5.2--5.3. Secondly, they can
avoid the need to produce {\it many\/} independent fine-level
configurations by making most of the sampling measurements at the coarse
levels of the algorithm: see Secs.\ 5.4--5.7. Thirdly, they can be used
to derive larger-scale equations of the model, thereby avoiding the need
to cover large domains by fine grids, eventually yielding macroscopic
equations for the model: see Sec.6.\par
This potential has so far been realized only for very simple model
problems. Extensions to more complex problems are not sure, and certainly
not straightforward. But one may note here that a similar situation
prevailed two decades ago in multigrid PDE solvers. It took years of
systematic research, indeed still going on today (and partly reported
herein), to extend the full model-problem efficiency to complicated real
life problems.\par
\section{Multigrid PDE Solvers}
The classical multi-scale method is the multigrid solver for discretized
partial differential equations (PDEs). The standard (or ``textbook")
multigrid efficiency is to solve the system of discretized equations in
{\it few\/} (less than 10) ``{\it minimal work units\/}",  each
equivalent to the amount of computational work (operations) involved in
{\it describing\/} the discrete equations; e.g., if the system is linear
and its discretized version is described by a matrix, this unit would be
the work involved in one matrix multiplication. (Note that even in case
of {\it dense\/} matrices, as in {\it integral\/} equations, multigrid
methods allow an $n\times n$ matrix multiplication to be performed in
only $O(n)$ or $O(n\log n)$ operations; see Sec.~6.)\par
Moreover, the multigrid algorithm can use a very high degree of {\it
parallel processing\/}: with enough processors it can, in principle,
solve a system of $n$ (discretized PDE) equations in only $O\bigl((\log
n)^2\bigr)$ steps. Note, for example, that each stage of the multigrid
cycles described above can be performed at all gridpoints in parallel;
only the stages themselves are sequential to each other.\par
The history of multigrid solvers is marked by systematic development,
which is gradually achieving the full standard efficiency stated above
for increasingly difficult and complex situations. Some highlights of
this development are presented below. For more details see the basic
guide \cite{A10}, with some general updates in \cite{A5}, \cite{A7}, and
more specific references given below. We will emphasize general features
that are relevant to the Dirac equations, but will postpone a specific
discussion of the latter to Sec.~4.\par
\subsection{Scalar linear elliptic equations}
An elliptic differential equation is often equivalent to a minimization
problem. For example, in $d$ dimensions ($x=(x_1,\ldots,x_d),\ \ptl_{\mu}
=\ptl/\ptl x_{\mu}$) the diffusion equation
\def\theequation{3.1}
\equa{\sum^d_{\mu=1}\ptl_{\mu}\bigl(a(x)\ptl_{\mu} U(x)\bigr)=f(x),\qquad
a(x)>0}
(with suitable boundary conditions) is equivalent to finding the function
$U$ which minimizes the energy
\def\theequation{3.2}
\equa{E(u)=\int a(x)\sum_{\mu}\bigl(\ptl_{\mu} u(x)\bigr)^2 dx}
among all functions $u$ for which this integral is well defined (and
which also satisfy suitable boundary conditions). The discretization of
(3.1) on a grid with meshsize $\alpha$ can also be formulated as a
minimization of a functional $E^{\alpha}(u^{\alpha})$. The multigrid $V$
cycle described above (Sec.\ 1.6) is a very efficient solver for such a
problem.\par
Instead of the formulation in terms of a minimization problem, the {\it
same\/} algorithm will now be written in terms of the PDE and its
discretization. This will enable us later to generalize it. Let the given
PDE, such as (3.1), be generally written as
\def\theequation{3.3}
\equa{Lu=f.}
Let the discrete solution at gridpoint $x_i^h$ of a given grid with
meshsize $h$ be denoted by $U_i^h$. The grid-$h$ approximation to (3.3)
can be written as
\def\theequation{3.4}
\equa{L^hU^h=f^h,}
where $L^h$ is an $n^h\times n^h$ symmetric and positive definite matrix.
Starting with any initial approximation $u^h$, a {\it multigrid cycle for
solving\/} (3.4) is recursively defined as the following 5 steps for
improving the approximation. (Note that $h$ here corresponds to $h/2$ in
Sec.\ 1.6.)\par
(i) {\it Pre-relaxation\/}. Perform $\nu_1$ relaxation sweeps on grid
$h$. In each relaxation sweep the gridpoints $x_1^h,x_2^h,\ldots,
x_{n^h}^h$ are scanned one by one. For each gridpoint $x_i^h$ in its
turn, the value $u_i^h$ (the current approximation to $U_i^h$) is
replaced by a new value for which equation (3.4) is satisfied (holding
$u_j^h$, for all $j\ne i$, fixed). This relaxation method is called {\it
Gauss-Seidel relaxation\/}, and is exactly equivalent to the
point-by-point minimization (cf.\ Sec.\ 1.5) of the energy
\def\theequation{3.5}
\equa{E^h(u^h)=(u^h,L^hu^h).}
\par\vskip -\parskip
(ii) {\it Coarsening\/}. The equations for the next coarser grid, grid
$2h$,
\def\theequation{3.6}
\equa{L^{2h}U^{2h}=R^{2h}}
should express the requirement that
\disp{E^{2h}(U^{2h})=\min E^{2h}(u^{2h}),}\noindent
where, similar to (1.15),
\disp{E^{2h}(u^{2h})=E^h(u^h+I_{2h}^hu^{2h}).}\noindent
Here $E^h$ is given by (3.5), $u^h$ is the current solution on grid $h$
(the final result of Step (i)), and $I_h^{2h}$ is the interpolation from
grid $2h$ to grid $h$, defined as in (1.13)--(1.14), by
\def\theequation{3.7}
\equa{(I_{2h}^h)_{i,k}=w_k^{2h}\biggl({x_i^h-x_k^{2h}\over 2h}\biggr).}
It is easy to see that this requirement is equivalent to defining
\def\theequation{3.8}
\equa{R^{2h}=I_h^{2h}(f^h-L^hu^h)}
and
\def\theequation{3.9}
\equa{L^{2h}=I_h^{2h}L^hI^h_{2h},}
where
\def\theequation{3.10}
\equa{I_h^{2h}=2^{-d}(I_{2h}^h)^T,}
superscript $T$ denoting the adjoint operator (transposition). The factor
$2^{-d}$ turns $I_h^{2h}$ into an averaging operator, so (3.8) means that
$R^{2h}$ is obtained by local averaging of the current {\it residual
field\/} $R^h=f^h-L^hu^h$. Also, with this factor, $L^{2h}$, defined by
(3.9), is a grid-$2h$ approximation to $L$; hence (3.6) is nothing but a
grid-$2h$ approximation to the error equation $L^he^h=R^h$, where $e^h=
U^h-u^h$ is indeed the error which $U^{2h}$ is designed to approximate.
In fact, instead of (3.8) any other averaging of $R^h$ will do, and
instead of (3.9) a simpler (and much cheaper computationally) $L^{2h}$
can usually be used, derived, e.g., by direct discretization of $L$ on
grid $2h$.\par
(iii) {\it Recursion\/}. If $2h$ is already the coarsest grid, solve
(3.6) directly (or iteratively; in any case this should be cheap, since
the number of unknowns is very small). Otherwise, perform $\gamma$
multigrid cycles for solving (3.6), starting with the trivial
approximation $u^{2h}=0$.\par
(iv) {\it Uncoarsening\/}. Interpolate $u^{2h}$ to grid $h$ and add it as
a correction to $u^h$; that is,
\def\theequation{3.11}
\equa{u^h\leftarrow u^h+I_{2h}^hu^{2h}.}
\par\vskip -\parskip
(v) {\it Post-relaxation\/}. Perform $\nu_2$ additional relaxation sweeps
on grid $h$.\quad $\blbox$\par
The multigrid cycles thus defined, for cycle indices $\gamma=1$ and
$\gamma=2$, are displayed in Fig.~1 above.\par
The same cycles can be used even in case the elliptic problem is {\it
not\/} equivalent to a minimization problem. Correspondingly, there is
much freedom in selecting the algorithm components (relaxation,
inter-grid transfers $I_{2h}^h$ and $I_h^{2h}$, and the coarse grid
operator $L^{2h}$) and in treating boundaries (see Sec.\ 3.2 below).\par
In fact, the efficiency of the multigrid cycle (e.g., its asymptotic
convergence factor) can exactly be predicted in advance by {\it local
mode\/} ({\it Fourier\/}) {\it analysis\/} \cite{A6}; so exactly indeed
that the prediction can be used in algorithmic design (choosing optimal
relaxation, inter-grid transfers, etc.) and program debugging. In
particular, it yields general rules (summarized in Sec.\ 3.4 below) for
the required orders of interpolation $I_{2h}^h$ and restriction
$I_h^{2h}$, and a general, easily computable yardstick for measuring the
efficiency of the interior (away from boundaries) relaxation, called the
{\it smoothing factor of relaxation\/}, $\ov{\mu}$ \cite{A9}. This
$\ov{\mu}$ is defined as the factor by which each sweep of relaxation
reduces the {\it high-frequency\/} components of the error, i.e., those
components that cannot be reduced by the coarse-grid corrections. With
proper choice of the inter-grid transfer, the error reduction factor per
multigrid cycle should approach $\ov{\mu}^{\nu_1+\nu_2}$.\par
For uniformly elliptic equations of {\it second order\/}, Gauss-Seidel
(GS) relaxation often yields the best smoothing factor per operation per
point. It is usually most effective when done in Red-Black (RB) ordering:
first the ``red" points (say those $x_j^h=(x_{j1}^h,\ldots,x_{jd}^h)$
with even $\sum_{\alpha} x_{j\alpha}^h/h$) are relaxed, then the ``black"
ones (with odd $\sum_{\alpha} x_{j\alpha}^h/h$). In the case of the
Poisson equation ((3.1) with constant $a(x)$) in two dimension $(d=2)$,
for example, RB-GS relaxation costs only 4 additions per point and yields
$\ov{\mu}=.25$ \cite{A10}. Relaxation of {\it higher order\/} uniformly
elliptic scalar equations, such as the biharmonics equation, can attain a
similar efficiency by writing them as a system of second order equations,
each relaxed by RB-GS. When the uniform ellipticity deteriorates, other
relaxation schemes should be employed: see Secs.\ 3.6 and 3.7 below, and
Sec.~3 in \cite{A10}.\par
How many cycles are needed to solve the problem (3.4)? This depends on
the quality of the initial approximation, $u^{h0}$, and on the required
size of the final error $\parallel u^h-U^h\parallel$. In particular, if
the following algorithm is used, only {\it one\/} cycle will often do.
\par
The {\it full-multigrid algorithm with\/ $n$ cycles per level\/}, or
briefly the $n$-{\it FMG Algorithm\/} for level $h$ is recursively
defined as follows. If $h$ is the coarsest grid, solve (3.4) directly.
Otherwise, apply first the $n$-FMG algorithm to the corresponding
equation on grid $2h$
\def\theequation{3.12}
\equa{L^{2h}U^{2h}=f^{2h}\eqd I_h^{2h}f^h,}
obtaining for it an approximate solution $u^{2h}$. Then interpolate the
latter to the fine grid
\def\theequation{3.13}
\equa{u^{h0}=\II_{2h}^hu^{2h},}
and perform $n$ multigrid cycles for level $h$ to improve this initial
approximation, yielding the final approximation $u^h$.\par
An example of the entire flow of the 1-FMG algorithm, employing one $V$
cycle at each level, is shown in Fig.~2. Such an algorithm (with $\nu_1+
\nu_2=2$ or 3) is usually enough to reduce the error below discretization
error, i.e., to yield
\begin{figure*}[htb]
\vspace*{5truecm}
\caption{{\it FMG Algorithm\/} with 4 levels and one $V$ cycle per level.
A crossed circle $\oplus$ stands for a direct solution of equation on a
coarsest grid. A double upward arrow ~~~~ indicates interpolation of
solution to a new level. All other notations are the same as in Fig.~1.}
\end{figure*}%
\def\theequation{3.14}
\equa{\parallel u^h-U^h\parallel\le C\parallel U^h-U\parallel,}
(with $C=.5$, say), {\it provided\/} the problem is uniformly elliptic,
and the proper relaxation scheme and inter-grid transfers are used. In
particular, the required order of the {\it solution\/} interpolation
(3.13) (unlike the order of the {\it correction\/} interpolation (3.11))
depends on the norm $\parallel\cdot\parallel$ for which one wants (3.14)
to be satisfied: see Sec.\ 3.4.\par
One can of course append additional $n_{\ast}$ cycles to the 1-FMG
algorithm, thus satisfying (3.14) with much smaller $C$; typically
$C\approx 10^{-n_{\ast}}$. This may be wasteful: it will bring $u^h$
closer to $U^h$, but not closer to the differential solution $U$.\par
\subsection{Various boundaries and boundary conditions}
Along with the interior equations (approximating the PDE), the
discretized boundary conditions should also be relaxed, and their
remaining residuals should be averaged and transfered to serve as the
forcing terms of the boundary conditions on the next coarser grids.\par
Special attention should be paid to the interior equations in the
vicinity of the boundary. The error-smoothing effect of relaxation is
disrupted there in various ways, and the correct representation of the
near-boundary residuals on the coarser grid is generally more complicated
than (3.8) and depends on the shape of the boundary and the type of
boundary conditions.\par
A general way around these difficulties is to add extra relaxation steps
near the boundary, especially near re-entrant corners and other
singularities. With the suitable relaxation scheme this can reduce the
near-boundary interior residuals so much that their correct
representation on the coarser grid is no longer important.\par
The work added by such near-boundary extra relaxation steps is negligible
compared to that of the full relaxation sweeps. It can be proved that
with such steps the efficiency of the multigrid cycle becomes independent
of the boundary shape and the type and data of the boundary conditions
\cite{A6}. This has been shown (experimentally) to be true even in
particularly difficult boundary situations, such as: highly oscillatory
boundary curves and/or boundary data and/or boundary operators (with the
oscillation wavelength comparable to the meshsize) \cite{A17}; free
boundaries \cite{A37}; ``thin" domains (much thinner than the meshsize of
some of the employed coarser grids) \cite{A36}; domains with small holes
(see Sec.\ 3.9 below); etc.\par
\subsection{FAS: nonlinear equations, adaptive resolution}
A very useful modification to the multigrid cycle is the {\it Full
Approximation Scheme\/} ({\it FAS\/} \cite{A9}), in which the coarse grid
equations (3.6) are rewritten in terms of the ``full approximation
function"
\def\theequation{3.15}
\equa{\ov{U}^{2h}=\ov{I}_h^{2h}u^h+U^{2h},}
where $\ov{I}_h^{2h}u^h$ is the representation on the coarse grid of
$u^h$, the current fine grid approximation. This yields the coarse-grid
equation
\def\theequation{3.16}
\equa{L^{2h}\ov{U}^{2h}=\ov{f}^{2h}\eqd L^{2h}(\ov{I}_h^{2h}u^h)+R^{2h}.}
In the uncoarsening step, $u^{2h}$ in (3.11) should of course be replaced
by $\ov{u}^{2h}-\ov{I}_h^{2h}u^h$, where $\ov{u}^{2h}$ is the approximate
solution to (3.16) obtained in the recursion step.\par
One advantage of the FAS is that it can directly be applied to {\it
nonlinear\/} $L^h$, without any linearizations: the same simple $L^{2h}$
can serve in (3.16) as in (3.12) (whereas in (3.6) it could not, unless
the equations are linear). Since $L^h,L^{2h},\ldots$ are all similar,
unified programming for all levels is facilitated.\par
The FAS-FMG algorithm solves nonlinear problems as fast as linear ones,
namely, in less than 10 minimal work units. No Newton-Raphson iterations
are required. Also, solution tracing processes (embedding, continuation,
searches in bifurcation diagrams, etc.), needed in case of severe
nonlinearities, can be performed very cheaply, by procedures which employ
the finest grid very rarely. For example, continuation processes can
often be integrated into the FMG algorithm, at no extra cost; i.e., a
problem parameter gradually advances as the solver (cf.\ Fig.~2) proceeds
to finer levels.\par
A general advantage of the FAS is that averages of the full solution, not
just corrections, are represented on all coarser grids (hence the name of
the scheme). This allows for various advanced techniques which use finer
grids very sparingly. For example, the fine grid may cover only part of
the domain: outside that part $\ov{f}^{2h}$ of (3.16) will simply be
replaced by the original coarse grid right-hand side, $f^{2h}$ of (3.12).
One can use progressively finer grids confined to increasingly more
specialized subdomains, effectively producing better resolution only
where needed. In this way an {\it adaptive\/} resolution is formulated in
terms of {\it uniform\/} grids, facilitating, e.g., low-cost high-order
discretizations as well as fast multigrid solvers. The grid adaptation
itself (i.e., deciding {\it where\/} to introduce the finer level) can be
done with negligible extra work (no repeated solutions) by being
integrated into the FMG algorithm (at the double-arrow stages of Fig.~2).
As a refinement {\it criterion\/} (indicating where a further refinement
of the currently-finest grid $h$ is {\it needed\/}) one can use the local
size of $\vert \ov{f}^{2h}-f^{2h}\vert$, which measures the correction
introduced by grid $h$ to the grid-$2h$ {\it equations\/}. Moreover, each
of the local refinement grids may use its own {\it local coordinate
system\/}, thus curving itself to fit boundaries, fronts,
discontinuities, etc. Since this curving is only local, it can be
accomplished by a trivial transformation, and it does not add substantial
complexity in the bulk of the domain (in contrast to global
transformation and grid generation techniques). See details in Secs.\ 7-9
of \cite{A9}, Sec.~9 of \cite{A10}, \cite{A1} and a somewhat modified
approach in \cite{A34}.\par
\subsection{Non-scalar PDE systems}
A system of $q$ differential equations in $q$ unknown functions is called
{\it non-scalar\/} if $q>1$. General multigrid procedures have been
developed for solving (the discretized version of) such systems. The
overall flow of the algorithm remains the same (Figs.~1 or 2). General
rules were developed (in \cite{A10}) for deriving suitable relaxation
schemes and inter-grid transfers for any given system.\par
The most important rules of inter-grid transfers are the following. (For
more details see \cite{A10} and \cite{A6}.) Let $m_{ij}$ denote the order
of differentiation of the $j$-th unknown function in the $i$-th
differential equation. Let $m^j$ denote the order of the {\it
correction\/} interpolation $I_{2h}^h$ applied in (3.11) to the $j$-th
unknown function, and $m_i$ the order of the fine-to-coarse transfer
$I_h^{2h}$ applied in (3.8) to the residuals of the $i$-th equation. (If
(3.10) is used then $m_i=m^i$. The order of interpolation is defined at
(1.8).) Let $M^j$ be the order of the {\it solution\/} interpolation
$\II_{2h}^h$ applied in (3.13) to the $j$-th function. Let $p$ denote the
order of discretization, i.e., $\parallel U^h-U\parallel=O(h^p)$.
Finally, let $\ell_j$ be the order of derivatives we want to calculate
for the $j$-th function, i.e., the order of its derivatives entering into
the norm $\parallel\cdot\parallel$ used in (3.14). Then, to guarantee the
full possible efficiency of the multigrid cycle it is required that
\def\theequation{3.17}
\equa{m_i+m^j>m_{ij}.}
In the border case $m_i+m^j=m_{ij}$ the algorithm {\it may\/} sometimes
still perform satisfactorily. To guarantee further that the minimal
number of cycles is used it is also required that
\def\theequation{3.18}
\equa{M^j\ge p+\ell_j.}
\par\vskip -\parskip
In the case of first order systems, such as Dirac equations, since $m_{i
j}=1$ it follows that $m_i=m^j=1$, $(i,j=1,\ldots,q)$. These minimal
orders are very convenient: they mean that in uncoarsening, each $\delta
u^{j,h}_{\ell}$ can simply be taken as the value of any neighboring
$u_k^{j,2h}$, and in coarsening the residual $R_{\ell}^{i,h}$ can be
added to any neighboring $R_k^{i,2h}$.\par
The main tool developed (in \cite{A10}) for analyzing and discretizing
general non-scalar schemes, and for deriving suitable relaxation schemes
for them, is the {\it principal determinant operator\/}. To illustrate
this tool and its uses, consider for example the Cauchy-Riemann system
\def\theequation{3.19a}
\equa{\ptl_1 U^1+\ptl_2 U^2=f^1}
\def\theequation{3.19b}
\equa{\ptl_2 U^1-\ptl_1 U^2=f^2}
where $\ptl_{\nu} U^{\mu}=\ptl U^{\mu}/\ptl x_{\nu}$. It can be written
in the matrix operator form
\def\theequation{3.20}
\equa{LU=f,}
where
\disp{L=\left(\matrix{\ptl_1&\ptl_2\cr \ptl_2&-\ptl_1\cr}\right),\quad
U=\left(\matrix{U^1\cr U^2\cr}\right),\quad f=\left(\matrix{f^1\cr
f^2\cr}\right).}\noindent
The principal determinant operator in this case is simply the determinant
of $L$
\def\theequation{3.21}
\equa{\det L=-\ptl_1^2-\ptl_2^2=-\Delta,}
i.e., the Laplacian.\par
For more complicated nonlinear systems of nonuniform order, one has to
include only the {\it principal\/} part of the determinant of the {\it
linearized\/} operator. The principal part depends on the scale under
consideration. At sufficiently small scales the principal part is simply
the terms of $\det L$ with the highest-order derivatives.\par
{}From (3.21) one can learn that, like the Laplace equation, the
Cauchy-Riemann system is second-order elliptic and requires {\it one\/}
boundary condition (although there are {\it two\/} unknown functions).
\par
Any discretized version of (3.19) can analogously be written in the form
\def\theequation{3.22}
\equa{L^hU^h=f^h,}
\disp{\begin{array}{l}\noindent
L^h=\left(\matrix{\ptl_1^h&\ptl_2^h\cr \ptl_2^h&-\ptl_1^h\cr}
\right),\quad U^h=\left(\matrix{U^{1,h}\cr U^{2,h}\cr}\right),\\
\noalign{\vskip 5pt}
f^h=\left(\matrix{f^{1,h}\cr f^{2,h}\cr}\right).
\end{array}}
For example, if conventional (non-staggered) central differencing is
used, then $\ptl_1^h$ is defined, for any function $\varphi$, by
\def\theequation{3.23}
\equa{\begin{array}{ll}
\ptl_1^h\varphi(x_1,x_2)=&{1\over 2h}\bigl\{\varphi(x_1+h,x_2)\\
&-\varphi(x_1-h,x_2)\bigr\},
\end{array}}
and similarly $\ptl_2^h$. In this case
\def\theequation{3.24}
\equa{\det L^h=-(\ptl_1^h)^2-(\ptl_2^h)^2=-\Delta^{2h},}
where $\Delta^{2h}$ is the usual 5-point discrete Laplacian, except that
it is based on intervals $2h$, {\it twice\/} the given meshsize. From
this one can learn that this central discretization of the Cauchy-Riemann
system suffers from the same troubles as $\Delta^{2h}$, namely:\par
(1) The given grid $h$ is decomposed into 4 disjoint subgrids, each with
meshsize $2h$. The equations on these subgrids are decoupled from each
other locally, thus forming 4 decoupled subsystems. Hence the discrete
solution will have large errors in approximating solution {\it
derivatives\/} (whenever the discrete derivative involves points from
different subgrids). Note that in the case of the Cauchy-Riemann
equations, each subsystem involves values of both $U^{1,h}$ and $U^{2,
h}$, but at {\it different locations\/}.\par
(2) The discretization is wasteful, since it obtains grid-$2h$ accuracy
with grid-$h$ labor.\par
(3) The usual multigrid cycle will run into the following difficulty. The
error components slow to converge by relaxation may be smooth (i.e.,
locally nearly a constant) on each subgrid, but generally they must be
highly oscillatory on the grid as a whole (the 4 local constants being
unrelated to each other). Such an oscillatory error cannot be
approximated on a coarser grid.\par
One way to deal with the latter trouble is to regard $U^h$ (or each
$U^{\mu,h}$, in the Cauchy-Riemann case) as a vector of 4 different
functions, one function on each subgrid, with a corresponding
decomposition of the Laplacian (or the Cauchy-Riemann operator) into 4
operators, locally disconnected from each other. With a similar
decompositions on coarser grids, the coarse-grid corrections to any
subgrid on level $h$ should only employ values from the {\it
corresponding\/} subgrid on level $2h$, and each fine grid residual
should only be transferred to {\it corresponding\/} coarse grid
equations. With such decompositions the multigrid cycle will restore its
usual efficiency.\par
Another way to treat the same trouble is to employ the usual cycle,
without decompositions, which would necessarily lead to slow asymptotic
convergence, but to observe that the slow-to-converge components are
those highly oscillatory components which are smooth on each subgrid.
Such components have no counterpart in the differential solution, so they
can simply be (nearly) eliminated by solution averaging \cite{Quasi}.\par
Still another way, for dealing actually with {\it all\/} the troubles
listed above, is to observe that any 3 of the 4 subsystems are redundant,
so they can simply be dropped, turning what we called $2h$ into the
actual meshsize of the remaining grid. In case of the Cauchy-Riemann
system this would give the {\it staggered\/} discretization shown in
Fig.~3.\par
\begin{figure*}[htb]
\vspace*{6truecm}
\caption{{\it Staggered discretization of the Cauchy-Riemann system\/}.
$\btri$ and $\bbox$ show the location of $U^{1,h}$ and $U^{2,h}$
respectively. The discretizations of (3.19a) and (3.19b) are centered at
grid vertices and plaquette centers respectively.}
\end{figure*}%
The design of an efficient relaxation scheme for any $q\times q$
non-scalar operator $L$ can always be reduced to the design of a scheme
for each of the factors of the scalar operator $\det L^h$. Namely,
writing
\disp{\det L^h=L_1^h\cdots L_s^h}\noindent
where each factor $L_j^h$ is a first- or second-order operator. Once a
relaxation scheme has been devised for each $L_j^h$, with a smoothing
factor $\ov{\mu}_j$, these $s$ schemes can be composed into a relaxation
scheme for the original system $L^h$, with a smoothing factor $\ov{\mu}=
\max_j\ov{\mu}_j$. See \cite{A10} for details. In particular, one can
relax the Cauchy-Riemann system, and many other elliptic systems for
which $\det L^h=\pm(\Delta^2-m^2)^s$, as efficiently as relaxing the
Laplacian, obtaining $\ov{\mu}=.25$ (cf.\ Sec.\ 3.1).\par
In the case of the Cauchy-Riemann and similar first-order systems, a
simpler (but less general) description of the same relaxation is that it
is a Kaczmarz relaxation scheme.\par
{\it Kaczmarz relaxation\/} \cite{Kc}, \cite{Tn}, \cite{AMGT},
\cite{A6}, for a general linear system of real or complex equations
\disp{\sum^n_{j=1} a_{ij}U_j=f_i,\qquad (i=1,\ldots,n),}\noindent
is defined as a sequence of steps, each one relaxing one of the
equations. Given an approximate solution $u=(u_1,\ldots,u_n)$, a {\it
Kaczmarz relaxation step for the\/ $i$-th equation\/} is defined as the
replacement of $u$ by the vector closest to it on the hyperplane of
solutions to the $i$-th equation. This means that each $u_k$ is replaced
by $u_k+\beta_i a^{\ast}_{ik}$, where $a^{\ast}_{ik}$ is the complex
conjugate of $a_{ik}$ and
\disp{\beta_i=\biggl(f_i-\sum^n_{j=1} a_{ij}u_j\biggr)/\sum^n_{j=1}\vert
a_{ij}\vert^2.}
Kaczmarz relaxation always converges to a solution, if one exists, but
the speed of convergence may be extremely slow. For scalar elliptic
systems its smoothing factors are much poorer than those obtained by
Gauss-Seidel. But for first order non-scalar systems it can attain
optimal smoothing factors. Kaczmarz relaxation for Cauchy-Riemann
equations in Red-Black ordering for each of the two equations (which
could be derived, as mentioned above, from RB-GS for the Laplacian), has
the smoothing factor $\ov{\mu}=.25$. In usual ordering (row by row or
column by column) the smoothing factor is only $\ov{\mu}=.5$.\par
\subsection{Discontinuous and disordered coefficients. AMG}
In the case of discontinuous or very non-smooth data, the multigrid
algorithm requires certain modifications to retain its full above-stated
``textbook" efficiency.\par
The discontinuity of the external field $f^h$ (cf.\ (3.4)) does not
affect the efficiency of the multigrid cycle. It only requires some
attention in the FMG algorithm: the discrete coarse-grid field $f^{2h}$
(cf.\ (3.12)) should be formed by a full averaging of $f^h$; e.g.,
$f^{2h}=I_h^{2h}f^h$ with $I_h^{2h}$ of the type (3.10). Also, when $f^h$
has a strong singularity (e.g., it is a delta function, representing a
source), some special relaxation passes should be done over a small
neighborhood of that singularity immediately following the solution
interpolation to a new level (the double arrow in Fig.~2).\par
More difficult is the case of discontinuities in the principal terms;
e.g., in the diffusion coefficient $a(x)$ (cf.\ (3.1) or (3.2)).
Experiments revealed that the convergence factor per multigrid cycle
completely deteriorates when $a(x)$ has strong discontinuities, in
particular when $a(x)$ discontinuously jumps from one size to an
orders-of-magnitude different size. The main difficulty, it turns out, is
to represent $a(x)$ correctly on the coarser grids. The conductance
(relatively large $a(x)$) and insulation (small $a(x)$) patterns can be
too finely complicated to be representable on coarse scales.\par
The first fairly successful approach to such strongly discontinuous
problems \cite{ABDP} was to employ the coarsening (3.6)--(3.10) with
special shape functions $w_k^{2h}$, such that, for each $x_i^h$, the
interpolation weight $w_k^{2h}\bigl((x_i^h-x_k^{2h})/2h\bigr)$ is
proportional to a ``properly averaged" (see below) value of the
transmission $a(x)$ between $x_i^h$ and $x_k^{2h}$. In fairly
complicated cases this creates $L^{2h}$ with quite faithful
conductance-insulation patterns, yielding fully efficient cycles. This
approach led to very useful ``black-box" solvers \cite{D1}.\par
Another possible approach is to derive $L^{2h}$ not by (3.9) but as a
direct discretization of $L$, with properly averaged values of $a(x)$.
``Proper averaging", as is well known in electrical networks, means {\it
harmonic\/} averaging (i.e., arithmetic averaging of the resistance
$a(x)^{-1}$) in the transmission direction and {\it arithmetic\/}
averaging in the perpendicular direction.\par
Both these approaches will fail in cases of conductance-insulation
geometry that cannot be approximated on a uniform coarse lattice. Coarse
levels should then abandon the lattice structure, adapting their geometry
to that of the problem.\par
In ``{\it algebraic multigrid\/}" (AMG) solvers, introduced in the early
1980's (\S13.1 in \cite{G82}, \cite{AMGA}, \cite{AMGS}, \cite{AMGT},
\cite{KS}, \cite{RS}), no grids are used. Even the spatial
geometry behind the {\it given\/} (the finest) algebraic system need not
be given explicitly (although its implicit existence may be important for
the sparsity of the coarser levels formed by the algorithm). The
next-coarse-level variables are typically selected by the requirement
that each current-fine-level variable is ``strongly connected" to at
least some coarse-level variables, the strength of coupling being
determined by the fine-level equations (e.g., by relative local sizes of
the discrete conductance). The inter-level transfers may also be purely
based on the algebraic equations, although geometrical information may be
helpful (see \cite{AMGT}, \cite{RS}).\par
AMG solvers usually involve much (one to two orders of magnitude) more
computational work, and also much more storage than the regular
(``geometric") multigrid. Also, they have been successfully developed so
far only for limited classes of problems (mostly scalar, and having
``local positive definiteness" \cite{AMGT}). However, they are often
convenient as black boxes, since they require no special attention to
boundaries, anisotropies and strong discontinuities, and no
well-organized grids (admitting, e.g., general-partition finite element
discretizations). More important, AMG solvers are indispensable for {\it
disordered systems\/}, such as diffusion problems with arbitrary
conductance-insulation patterns, or problems not derived from a PDE at
all, such as the geodetic problem (which motivated much of the original
AMG work \cite{AMGA}), the random-resistor network (which served to
introduce AMG to the world of physicists \cite{AS16}), the Laplace
equation on random surfaces \cite{BA1}, and many others.\par
Also, AMG-type solvers can be developed for new classes of problems. See
the discussion in Sec.\ 4.6 below.\par
\subsection{Anisotropic equations. Convection dominated flows}
Good ellipticity measures (e.g., domination of viscosity) at all scales
of the problem is essential for the success of the multigrid algorithms
described above, since ellipticity means that non-smooth solution
components can be calculated by purely local processing. (See Sec.~2 of
\cite{A10} for general definitions of ellipticity measures.)\par
Small-ellipticity problems are marked either by indefiniteness ---
discussed in Sec.\ 3.7 below --- or by anisotropies. In the latter case,
characteristic directions (directions of strong coupling, or convection
directions, etc.) exist in the equations. Non-smooth solution components
can be convected along the characteristics, hence they cannot be
determined locally. Therefore, to still obtain the ``textbook" efficiency
stated above (beginning of Sec.~3), the multigrid algorithm must be
modified.\par
Fully efficient multigrid algorithms have been developed by using the
characteristic directions in various ways; e.g., in devising the
relaxation scheme. See Sec.\ 3.3 in \cite{A10} for the case that the
characteristics are aligned with gridlines, and \cite{A4}, \cite{A19},
\cite{A20}, \cite{A21} for the non-aligned case.\par
Most multigrid codes in use today for high-Reynolds (small viscosity)
steady-state flows do not incorporate this type of modifications.
Therefore, although yielding large improvements over previous one-grid
solvers, they are very far from attaining the ``textbook" multigrid
efficiency.\par
\subsection{Indefinite problems. Wave equations}
Indefinite problems arise as the spatial part of wave equations in
acoustics, seismology, electromagnetic waves and quantum mechanics. A
model example is the real equation
\def\theequation{3.25}
\equa{\Delta U(x)+k(x)^2U(x)=f(x).}
\par\vskip -\parskip
{\it Slight indefiniteness\/}. If $k(x)^2$ is everywhere small, so that
only few eigenvalues of (3.25) are positive, the algorithm does not
change on fine levels (except when some eigenvalues come too close to 0;
see Sec.\ 3.9). But at a certain coarse grid one can no longer solve fast
by using still coarser grids. This grid is very coarse, though, since it
only has to provide approximations to those (very smooth) eigenfunctions
corresponding to the positive eigenvalues, hence one can efficiently
solve there, e.g., by Gaussian elimination or Kaczmarz relaxation (see
end of Sec.\ 3.4) \cite{NSSI}.\par
If $k(x)^2$ is generally small, still making only few eigenvalues
positive, but it is large in some small regions (creating {\it local
indefiniteness\/}), the same algorithm applies, except that Kaczmarz
relaxation should be used at those special regions on {\it all\/} grids.
Since the Kaczmarz smoothing is poor, more relaxation passes should be
made over those regions.\par
{\it High indefiniteness\/} is the more difficult case when the
wavelength $\lambda(x)=2\pi/k(x)$ is generally small compared to the
linear size of the problem domain. Oscillatory solution components with
wavelength near $\lambda(x)$ are not determined locally. Hence, on scales
at which such components are non-smooth --- i.e., on grids with meshsize
$h$ comparable to $\lambda(x)$ --- the multigrid solver must be radically
modified.\par
The modified approach, similar to that which has been developed for
integral equations with oscillatory kernels \cite{A8}, is to represent
the solution on each level $h$ as a sum
\disp{U^h(x)=\sum^{m^h}_{j=1} A_j^h(x)e^{ik(x)\xi_j^h\cdot x}}\noindent
where the $\xi_j^h$ are $d$-dimensional unit-length vectors, uniformly
covering the unit sphere, their number being $m^h=O(h^{1-d})$. Thus, for
increasingly coarser spatial grids, increasingly finer momentum
resolution (denser grids $\xi_j^h$) are used. In some version of this
approach, on sufficiently coarse levels the equations will become similar
to ray formulations (geometrical optics).\par
This approach can yield not only fast solvers to discretized standing
wave equations, such as (3.25), but also the option to treat most of the
problem domain on coarse levels, hence essentially by geometrical optics,
with nested local refinements (implemented as in Sec.\ 3.3 above)
confined to small regions where the ray formulation breaks down. Also,
the implementation of radiation boundary conditions in this approach is
straightforward.\par
Such representations can also form the basis for very efficient multigrid
algorithms to calculate {\it many eigenfunctions\/} of a given elliptic
operator; e.g., the Schr\"odinger operator in condensed matter
applications.\par
\subsection{Small-scale essential features}
A small-scale essential feature is a feature in the problem whose linear
dimension is comparable to the meshsize $h$ but whose influence on the
solution is crucial. Examples: a small hole in the domain (an island), on
whose boundary the solution is prescribed; a small piece of a boundary
where a different type of boundary condition is given; a small but deep
potential well or potential barrier in the Schr\"odinger equation; large
``topological charge" over few plaquettes in Dirac equations; etc. When
its linear size is too small, such an essential feature may become
invisible to the next coarser grid $2h$, yielding wrong coarse-grid
approximations to smooth errors.\par
One general way around this difficulty is to {\it enlarge\/} the feature
so that it becomes visible to grid $2h$, then enlarge it again on
transition to grid $4h$, and so on. The enlarged feature should be
defined (e.g., the depth of the enlarged potential well should be chosen)
so that its global effect remains approximately the same. Also, to make
up for the local mismatch, the uncoarsening step (see Sec.\ 3.1) should
be followed by special relaxation steps in the vicinity of the small
features on the fine grid.\par
If there are many, or even just several, small-scale essential features
in the problem, then on a certain coarse grid they become so crowded
locally that they can no longer be further enlarged. On such a grid,
however, a proper relaxation scheme can usually provide a fast solver,
without using still coarser grids at all. This approach was successfully
implemented in the case of small islands \cite{VPH}.\par
Another general approach is to ignore the small feature on grid $2h$
altogether. This would normally cause the multigrid algorithm to slow
down, sometimes even to diverge. A closer look shows however that most
solution components still converge fast; only few, say $\nu$,
well-defined components do not. The slow to converge error can therefore
be eliminated by {\it recombining\/} $\nu+1$ iterants of the algorithm.
This means replacing the configuration obtained at the end of each cycle
by a linear combination of the configurations obtained at the end of the
last $\nu+1$ cycles, so as to obtain the least square norm of the
residuals. Here again, local relaxation steps in the vicinity of the
small feature should be added. This method has been tried for a variety
of small-scale features and was found to restore the full ``textbook"
multigrid efficiency \cite{A16}.\par
\subsection{Nearly singular equations}
An alternate explanation for the efficiency of the usual multigrid cycle
for an elliptic equation on any grid $h$ can be given in terms of
eigenfunctions, as follows. An eigenfunction amplitude fails to converge
efficiently in relaxation only if the corresponding eigenvalue,
$\lambda^h$, is in magnitude much smaller than other eigenvalues on grid
$h$. Such an eigenfunction, however, is so smooth that it is well
approximated on the coarse grid; that is, the corresponding eigenvalue on
the coarse grid, $\lambda^{2h}$, satisfies
\def\theequation{3.26}
\equa{\vert\lambda^h-\lambda^{2h}\vert\ll\vert\lambda^{2h}\vert,}
which guarantees a small relative error in the coarse-grid approximation
to the eigenfunction.\par
We call an equation ``nearly singular" for grid $h$ if some of its
smallest eigenvalues $\lambda$ are so small that the discretization error
in $\lambda$ (on grid $h$ or $2h$ or both) is comparable to $\lambda$, so
that (3.26) fails. The troublesome eigenfunctions are called ``almost
zero modes" (AZMs).\par
To obtain the usual multigrid efficiency such modes should be eliminated.
This can be done by recombining iterants, as in Sec.\ 3.8 above. See also
in \cite{NSSI}, \cite{A16}.\par
\subsection{Time dependent problems and inverse problems}
For parabolic {\it time-dependent problems\/} it has been shown that
multigrid techniques are extremely efficient not just in that they solve
fast the system of implicit equations at each time step \cite{A14}. A
large additional benefit is that only rare activation of fine scales is
needed wherever the solution changes smoothly in time; e.g., wherever and
whenever the forcing terms are stationary \cite{A24}. Also, multileveling
allows parallel processing not only at each time step, but across the
entire space-time domain. Extensions of such ideas to other time
dependent problems, including high-Reynolds flows, are currently under
study.\par
{\it Inverse problems\/} can become well-posed when formulated in a
multi-scale setting, and can be solved at a cost comparable to that of
solving corresponding direct problems \cite{A42}, \cite{A43}. A
demonstration of this is being developed for system identification and
inverse gravimetric problems.\par
\section{Multigrid Dirac Solver}
A major part of lattice field calculations is the inversion of the
discretized Dirac operator $L^h$ appearing in the fermionic action. This
is needed both by itself and also as part of calculations of the
determinant of $L^h$ in case of interacting fermions (see Sec.\ 4.9). For
this purpose, repeated solutions of systems of the type
\def\theequation{4.1}
\equa{L^h\Psi^h=f^h}
is needed. Multigrid solvers for this type of equations will be described
here. In addition, the multigrid solver can save most of the work in
repeatedly {\it re\/}-solving (4.1) for new gauge fields and forcing
terms $f^h$ (see Sec.\ 4.8). Multigrid fast gauge fixing will also be
described (Sec.\ 4.2).\par
The matrix $L^h$ itself depends on the lattice-$h$ gauge field $U^h$.
Note that the unknown function in (4.1) is therefore denoted by $\Psi^h$
here (instead of $U^h$ in the previous chapter), and its computed
approximations will be denoted $\psi^h$ (instead of $u^h$). Also,
$\Psi^h$ and $\psi^h$ here are {\it complex\/}, not real, functions. For
simplicity we will omit the superscript $h$ until we need to distinguish
between several levels of the algorithm.\par
The methods described here were developed at the Weizmann Institute in
collaboration with others: see \cite{AZM}, \cite{A2}, \cite{PT2},
\cite{PhDR}. Some of the reported ideas were developed in parallel by
other groups \cite{AS124}, \cite{AS125}.\par
\subsection{Model case: 2-D QED}
The main difficulties in solving lattice Dirac equations are already
exhibited in the simple case of the Schwinger model \cite{PT9} (two
dimensional QED). In staggered formulation \cite{PT11}, on fine enough
grid with meshsize $h$, equation (4.1) at gridpoint $(j,k)$ has the form
\def\theequation{4.2}
\equa{\begin{array}{ll}
L^h\Psi_{j,k}\eqd&D_1^h\Psi_{j,k}+(-1)^jD_2^h\Psi_{j,k}\\
&+m_q\Psi_{j,k}=f_{j,k},
\end{array}}
where $m_q$ is the quark mass and $D_{\mu}^h$ are the discrete
``covariant derivatives", defined by
\disp{D_1^h\psi_{j,k}={1\over 2h}\bigl(U^{\ast}_{j+1/2,k}\psi_{j+1,k}-U_{
j-1/2,k}\psi_{j-1,k}\bigr)}\noindent
\disp{D_2^h\psi_{j,k}={1\over 2h}\bigl(U^{\ast}_{j,k+1/2}\psi_{j,k+1}-U_{
j,k-1/2}\psi_{j,k-1}\bigr).}\noindent
Thus, the gauge field $U=U^h$ is defined on grid {\it links\/}. Each
value of $U$ is a complex number of magnitude 1, and $U^{\ast}$ is its
complex conjugate (hence inverse); i.e.,
\def\theequation{4.3}
\equa{U_{\ell}=e^{ihA_{\ell}},\qquad U^{\ast}_{\ell}=e^{-ihA_{\ell}},}
where $\ell=(j+1/2,k)$ or $(j,k+1/2)$ and $A_{\ell}$, the gauge field
{\it phase per unit length\/}, is {\it real\/}. Note the meshsize
dependence introduced in (4.3). Customarily, the finest-grid (the given
lattice) meshsize is $h=1$, but we do not confine our discussion to this
case since we will need coarser grids as well, and also since we will
like to discuss the limit $h\rightarrow 0$. (See the corresponding
differential equation, and an alternative discretization, in Sec.\ 4.7).
\par
{\it Gauge fluctuations\/}. We assume physically realistic gauge fields,
as produced, e.g., by the quenched approximation \cite{PT12}. This means
statistical fluctuations of $U$ according to the {\it gauge action\/}
\def\theequation{4.4}
\equa{S_G=\beta\Sigma\bigl[1-\cos(h^2\curl A_{j+1/2,k+1/2})\bigr],}
summation being over all plaquettes $(j+1/2,k+1/2)$ and the discrete curl
operator being defined by
\disp{\begin{array}{ll}
\curl A_{j+1/2,k+1/2}=&{1\over h}(A_{j+1/2,k}+A_{j+1,k+1/2}\\
&-A_{j+1/2,k+1}-A_{j,k+1/2}).
\end{array}}\noindent
This implies that at each plaquette $h^2\curl A$ has nearly Gaussian
distribution with mean 0 and variance $\beta^{-1}$.\par
{\it Gauge freedom\/}. The physical problem is unchanged (since so are
(4.2) and (4.4)) by any ``{\it gauge transformation\/}" of the form
\def\theequation{4.5a}
\equa{\psi_{j,k}\leftarrow\psi_{j,k}e^{iB_{j,k}},}
\def\theequation{4.5b}
\equa{U_{j+1/2,k}\leftarrow U_{j+1/2,k}e^{i(B_{j+1,k}-B_{j,k})}}
\def\theequation{4.5c}
\equa{U_{j,k+1/2}\leftarrow U_{j,k+1/2}e^{i(B_{j,k+1}-B_{j,k})}}
done at all sites and links, with any real grid function $B_{j,k}$. An
arbitrary such transformation can always be done on the problem.\par
{\it Correlation lengths\/}. Because of this freedom, the gauge
configuration may seem completely disordered. However, it is easy to see
that the integral of the field $A$ around any domain of area $\xi^2$,
thus containing $\xi^2/h^2$ plaquettes,has variance $\xi^2h^{-2}\beta^{-
1}$, hence the field $U$ has the correlation length $\xi_{_G}=O(\beta^{
1/2}h)$ (cf.\ Sec.\ 4.2). Other important lengths in the problem are the
matter correlation length $\xi_m=O(m_q^{-1})$, the pion correlation
length $\xi_{\pi}=O(\xi_m^{1/2}h^{1/2})$, and $Lh$, the linear size of
the lattice.\par
{\it Boundary conditions\/}. The true physical problems are given in the
entire space (or plane, in this case). The computations are done on a
finite $L\times L$ grid with {\it periodic\/} boundary conditions, which
may best approximate the unbounded domain. The periodicity of $U$
introduces some artificial topological difficulties (cf.\ Sec.\ 4.2),
and other boundary conditions would in fact be computationally easier.
\par
\subsection{Fast gauge fixing and updating}
To avoid the disorder introduced by the gauge freedom, one {\it can\/}
(although we will see that perhaps one does not {\it have to\/}) ``fix
the gauge" into a smooth field by a gauge transformation (4.5). To see
the smoothness, the field $A=A^h$ should first be recognized as a pair
of fields, $A^{1,h}$ and $A^{2,h}$, the first consisting of the values of
$A^h$ on horizontal links $\bigl(j+{1\over 2},k\bigr)$, the second on
vertical links $\bigl(j,k+{1\over 2}\bigr)$. One main reason for fixing
the gauge is to see that each fixed $A^{\mu,h}$ tends to a continuous
field $A^{\mu}$ as $h\rightarrow 0$, enabling a better understanding of
the fields, the equations, and the solver (cf.\ Sec.\ 4.7). The $2\pi$
periodicity of $hA^h$ is not meaningful at that limit, so we will fix the
gauge as a {\it real\/} field, loosing this periodicity. This is
impossible to do with periodic boundary conditions. (Indeed, the boundary
values are periodic in $U^h$, hence in $hA(\hbox{mod}\ 2\pi)$, not in $A$
itself.) Therefore we will fix the gauge {\it piecewise\/}, in a
rectangular subdomain $S$.\par
The sumdomain $S$ can be large, in fact as large as the entire domain,
but without its periodicity. Indeed we can even retain the periodicity in
one direction, $x_1$ say, and keep the full size of the domain in the
other direction, too, but with the periodicity cut out, say along the
gridline $k=k_0$. The simplest description will be in terms of a {\it
double value\/} for each $A^1_{j+1/2,k_0}$, denoted $A^1_{j+1/2,k_0-}$
and $A^1_{j+1/2,k_0+}$, referring to the sides $k<k_0$ and $k>k_0$
respectively. An actual gauge fixing will of course use only the values
of one side, $A^1_{j+1/2,k_0-}$ for example, calling them actually
$A^1_{j+1/2,k_0}$, giving for each $A^2_{j,k_0+1/2}$ a different value
than in our double-value description. Such a subdomain $S$ will be called
the {\it cut-torus cylinder\/}.\par
The values of $\vert h^2\curl A\vert$ determined (stochastically) by the
action (4.4) are not necessarily small even at large $\beta$; they are
only small {\it modulo\/} $2\pi$. Our first step is to turn them {\it
actually\/} small (not just modulo $2\pi$) throughout $S$, by adding to
$hA^1$, wherever needed, an integral multiple of $2\pi$. Note that we can
do that even in the cut-torus cylinder case, due to the permitted double
values along $k=k_0$. In fact, this operation is where double values are
being introduced.\par
{\it Note 1\/}.~~After this operation one can still add any arbitrary
(but the same) integer multiple of $2\pi$ to all $hA^1_{j+1/2,k}$ with
the same fixed $j$, and similarly to all $hA^2_{j,k+1/2}$ with the same
$k$.\par
Thus we get
\def\theequation{4.6}
\equa{\curl A=g,\qquad\max\vert g\vert<\pi h^{-2}}
prescribed by the action. The gauge freedom implies that the field
\disp{\begin{array}{ll}
(\hbox{div} A)_{j,k}=&{1\over h}(A_{j+1/2,k}-A_{j-1/2,k}\\
&+A_{j,k+1/2}-A_{j,k-1/2})
\end{array}}\noindent
can be fixed arbitrarily at all grid vertices $(j,k)$ --- at least on any
non-periodic piece of the lattice. To obtain as smooth a field as
possible, the {\it Landau gauge}
\def\theequation{4.7}
\equa{\hbox{div} A=0}
is natural. With the given near Gaussian fluctuations of $h^2\curl A$ on
each plaquettes, (4.7) implies that each of the fields $U^{h,\mu}=\exp(ih
A^{\mu})$ has correlation length $\xi_{_G}=O(\beta^{1/2}h)$; that is,
$\vert\xi A_{\ell}^{\mu}-\xi A_{\ell^{\prime}}^{\mu}\vert\ll\pi$ if the
distance $\xi$ between links $\ell$ and $\ell^{\prime}$ is small compared
with $\xi_{_G}$.\par
{\it Fast multigrid solver\/}. To explicitly find the field $B$ defining
the gauge transformation (2.5) that would yield (4.7) is equivalent to
solving a discrete 5-point Poisson equation for $B$. The FMG solver
(Fig.~2 above), employing RB-GS relaxation and $\nu_1=\nu_2=1$ would
solve the problem in less than 30 additions per gridpoint. (The needed
multiplications are by powers of 2, which can be performed as additions.)
\par
As discussed in Sec.\ 3.1, the FMG solver would give us solutions
satisfying something like (3.14). In the present case this means that any
better accuracy in solving (4.7) is not needed because it will not give
smaller variations in $A$, {\it on any scale\/}.\par
{\it Cut-torus gauge fixing\/}. A particularly useful gauge fixing can be
obtained for the cut torus cylinder. On the line $k=k_0$ double values of
the transformation field $B$ are allowed: $B_{j,k_0-}$ affecting $A^1_{j
\pm 1/2,k_0-}$ and $A^2_{j,k-1/2}$, and $B_{j,k_0+}$ affecting $A^1_{j\pm
1/2,k_0+}$ and $A^2_{j,k+1/2}$. We can require the gauge transformation
to give us
\def\theequation{4.8}
\equa{A^1_{j,k_0+}-A^1_{j,k_0-}=C_{\ast},\qquad\hbox{for all}\ j,}
where the constant $C_{\ast}$ is of course determined by the current
value of $\sum_j(A^1_{j,k_0+}-A^1_{j,k_0-})$, which has itself been
determined by the sum over all the plaquettes of the function $g$,
defined at (4.6); it is easy to see that $C_{\ast}hL/2\pi$ must be an
integer. The requirement (4.8) will determine $B_{j,k_0+}-B_{j,k_0-}$, so
only one of the two values remains at our disposal, and we denote it
$B_{j,k_0}$. We can now further require the gauge transformtion to yield
(4.7) {\it throughout the periodic domain, including the cutline\/}. (In
defining $\hbox{div}\ A$ at the cut point $(j,k_0)$ we can use for
$A^1_{j\pm 1/2,k_0}$ either $A^1_{j\pm 1/2,k_0+}$ or $A^1_{j\pm 1/2,k_0
-}$; due to (4.8) the result is the same.) This will give us again a
discrete Poisson equation for $B$, but with completely periodic boundary
conditions. The equation is solvable since the sum over its right-hand
sides vanishes. The undetermined additive constant in $B$ is immaterial
for the transformed field $A$. The fast multigrid solver described above
still applies; in fact, for these periodic boundary conditions it is
particularly simple, since no special treatment is needed at the
boundary. (For these periodic boundary conditions, and for convenient
lattice sizes such as $L=2^{\ell}$, one can also use fast FFT Poisson
solver, which is only slightly less efficient than multigrid; but it
would not have the super-fast updates described below.)\par
The result, which will be called the {\it cut-torus gauge field\/}, has
the field $A^2$ smooth {\it everywhere\/} (including at the cut), and the
field $A^1$ smooth everywhere except for a {\it constant\/} jump (4.8) at
the cut line $k=k_0$. This field is uniquely determined by (4.6), (4.7)
and the choice of $k_0$, except for the possible addition of constants
$2m_1\pi(hL)^{-1}$ and $2m_2\pi(hL)^{-1}$ to the fields $A^1$ and $A^2$
respectively, with arbitrary integers $m_1$ and $m_2$. This freedom
results from Note~1 above.\par
{\it Shifting the cut\/} from $k_0$ to $\ov{k}_0$ is a trivial
transformation. If for example $1\le k_0<\ov{k}_0\le L$, the new gauge
field $\ov{A}$ is, for all $j$,
\disp{\begin{array}{ll}
\ov{A}^1_{j+1/2,k}&=A^1_{j+1/2,k}+C_{\ast}+2m_1\pi(hL)^{-1},\\
&\quad (k_0<k<\ov{k}_0)\\
\ov{A}_{j+1/2,k_0}^1&=A^1_{j+1/2,k_0+}+2m_1\pi(hL)^{-1}\\
\ov{A}_{j+1/2,\ov{k}_0+}^1&=A^1_{j+1/2,\ov{k}_0}+C_{\ast}+2m_1\pi(hL)^{-
1}\\
\ov{A}_{j+1/2,\ov{k}_0-}^1&=A^1_{j+1/2,\ov{k}_0}+2m_1\pi(hL)^{-1}\\
\ov{A}_{j+1/2,k}^1&=A^1_{j+1/2,k}+2m_1\pi(hL)^{-1},\\
&\quad (k<k_0\ \hbox{or}\ k>\ov{k}_0),\\
\ov{A}^2&\equiv A^2+2m_2\pi(hL)^{-1},
\end{array}}\noindent
with an arbitrary choice of the integers $m_1$ and $m_2$. Saying that
{\it the cut is shifted away from a certain neighborhood\/} will
generally mean a choice of $k_0$, $m_1$ and $m_2$ so that $\ov{A}$ is not
only smooth but also as small as possible in that neighborhood.\par
Thus, the cut-torus gauge has the advantage that it immediately
describes, upto this simple shift, a field which is smooth at any chosen
neighborhood. We will see its use in Sec.\ 4.7 below.\par
{\it Super-fast updates\/}. Suppose the gauge field has been fixed as
described above, and then one of its values has been stochastically
changed. To {\it re\/}-fix the gauge it is enough to apply local
relaxation near the changed value. Only once in several such changes on
grid $h$ one needs to go {\it locally\/} to the coarser grid $2h$:
transfer {\it local\/} residuals to grid $2h$, as in (3.8), then relax
{\it locally\/} on the grid-$2h$ equations and correct the solution on
grid $h$ as in (3.11). Once in several such transfers to grid $2h$, a
similar transfer is made from grid $2h$ to grid $4h$; etc. In this way
{\it the cost of fixing the gauge is only\/ $O(1)$ per update, and the
fields\/ $A^1$ and\/ $A^2$ are kept as smooth as they can be, on all
scales\/}.\par
The range of the local relaxation may be minimized by applying the
stochastic changes in a special {\it distributive\/} manner (cf.\ Sec.\
4.8). In particular, if instead of changing an individual phase at a time
one changes simultaneously the four phases of a plaquette $A_{j+1/2,k}$,
$A_{j+1,k+1/2}$, $A_{j+1/2,k+1}$ and $A_{j,k+1/2}$, by the amounts
$+\delta$, $+\delta$, $-\delta$ and $-\delta$ respectively, where
$\delta$ is determined stochastically by the action, then no re-fixing of
the field is needed {\it at all\/}: if (4.7) is satisfied before such a
change, it will continue to be satisfied after it.\par
A similar (piecewise or cut-torus) gauge fixing applies in any dimension
and for non-Abelian gauge fields.\par
\subsection{Vacuum gauge\/: $\xi_{_G}=\infty$}
Consider first the case $A\equiv 0$ and $m_q=0$. Eq.\ (4.2) is then
identical to (3.22). As discussed there, it would be decomposed into 4
Cauchy-Riemann subsystems decoupled from each other, except that, due to
the staggering in (4.2), only two such subsystems are present. Denoting
by $\Psi^1$, $\Psi^2$, $\Psi^3$ and $\Psi^4$ the function $\Psi$ at
gridpoints $(j,k)$ with ($j$ odd, $k$ even), ($j$ even, $k$ odd), ($j$
odd, $k$ odd) and ($j$ even, $k$ even) respectively, one subsystem
couples $\Psi^1$ with $\Psi^2$ and the other couples $\Psi^3$ with
$\Psi^4$. A multigrid solver can thus be built along the lines explain
in Sec.\ 3.4, yielding the standard multigrid efficiency.\par
Note in particular that corresponding to the four species of $\Psi$ there
are four kinds of equations, each one centered at the gridpoints of
another species. Similar four species and four kinds of equations are
defined on the coarse grid. Corrections should be interpolated to each
species from the corresponding species on the coarse grid. Similarly,
residuals of one kind of fine-grid equations should be transferred to the
same kind on the coarse grid.\par
{\it A comment on the doubling effect\/}. Even with an arbitrary gauge
field, for $m_q=0$ the above two subsystems remain decoupled. For $m_q\ne
0$, since supposedly $h\ll\xi_m$, the two subsystems are still only
weakly coupled locally. Such ``doubled states" do not correspond to a
physical reality. They can be removed by the non-staggered Wilson
discretization, which breaks chiral symmetry. Another, perhaps simpler
way to remove them (also breaking chiral symmetry, but keeping {\it
second-order accuracy\/}) is to take only one of the staggered subsystems
(e.g., $\Psi^1$ and $\Psi^2$) and replace the term $m_q\Psi_{j,k}$ in
(4.2) by
\def\theequation{4.9}
\equa{{m_q\over 2}(U^{\ast}_{j,k+1/2}\Psi_{j,k+1}+U_{j,k-1/2}\Psi_{j,k-1}
).}
The doubling effect thus removed, this discretization is more convenient
for developing the fast multigrid solver, and cost only half the solution
time. We emphasize however that, as in Sec.\ 3.4, the original system can
be solved just as fast (per unknown), even though less conveniently.\par
The case $A\equiv 0$ and $m_q\ne 0$ is solved basically by the same
algorithm. On grids with meshsize $h\ll\xi_m$, the principal local
operator is as before, so the same relaxation would have nearly the same
smoothing rate. On grids with $h\ge O(\xi_m)$ the Kaczmarz relaxation
will have no slowing down, so no grid coarser than that need be employed.
Standard multigrid efficiency is still easily obtained.\par
Applying the gauge transformation (4.5) to the $A\equiv 0$ case does not
change the problem; it only expresses it in new variables, explicitly
related to the old ones. Therefore, the solver need not change either, it
should only be expressed in terms of the new variables. This immediately
yields the following algorithm, for any {\it vacuum\/} (transformable to
$A\equiv 0$) gauge field:\par
1. The overall flow of the algorithm is still the same (e.g., Fig.~2; the
coarsest grid, as explained above, should have $O(\xi_m)$ meshsize,
unless $\xi_m\ge Lh$).\par
2. Relaxation is still Kaczmarz (see end of Sec.\ 3.4).\par
3. The intergrid transfers $I_{2h}^h$ and $I_h^{2h}$ should use the gauge
field as {\it parallel transporter\/}. This means that if a quantity is
transferred from site $x$ to site $y$ (e.g., a residual $R_x^h$
calculated on the fine grid is transferred to a coarse gridpoint $y$, or
a correction $U_x^{2h}$ calculated on the coarse grid is transferred to a
fine gridpoint $y$ as part of an interpolation), the quantity should be
multiplied by $U_1U_2\cdots U_m$, where $U_1,U_2,\ldots,U_m$ are the
gauge field values along a sequence of lattice links leading from $x$ to
$y$, taking of coarse $U^{\ast}_{\ell}$ instead of $U_{\ell}$ when the
link $\ell$ leads in the negative direction. It is easy to check that,
since $\curl A=0$, the transporter $U_1U_2\cdots U_m$ does not depends on
the chosen route from $x$ to $y$.\par
4. The coarse grid operator $L^{2h}$ is the exact coarse-grid analog of
(4.2), with the coarse-grid gauge field $A^{2h}$ obtained from the fine
grid $A^h$ by {\it injection\/}. Namely, if the coarse-grid link $L$ is
the union of the fine-grid links $\ell$ and $\ell^{\prime}$, then $A_L^{2
h}=(A_{\ell}^h+A_{\ell^{\prime}}^h)/2$, hence $U_L^{2h}=U_{\ell}^hU_{
\ell}^h$ (noting the dependence on $h$ introduced in (4.3)).\par
Step by step, this algorithm will record results (e.g., magnitude of
residuals) that are independent of the gauge transformation. Thus it will
still exhibit the textbook multigrid efficiency.\par
\subsection{Scales\/ $h\ll\xi_{_G}$ or\/ $\xi_{_G}\gsim\min(\xi_m,hL)$}
The case discussed above is that of $h^2\curl A=0(\hbox{mod}\ 2\pi)$,
produced for example by $\beta\rightarrow\infty$, giving also $\xi_{_G}
\rightarrow\infty$. The same algorithm can still be employed, and usually
at the same efficiency, on all grids with meshsize $h\ll\xi_{_G}$. On
such grids the parallel transporter is still locally well defined (nearly
route independent), hence the algorithm will perform locally close to its
performance for $A=0$.\par
By saying that a multigrid algorithm works efficiently {\it for a certain
meshsize\/} $h$ we mean the qualification ``provided the equation for
level $2h$ are solved efficiently; whether or not this provision holds is
a separate discussion".\par
Still, even on levels with $h\ll\xi_{_G}$, a certain trouble may arise;
in fact sometimes it does, sometimes it does not: the system of equations
may become {\it nearly singular\/} (cf.\ Sec.\ 3.9). This for example
happens when $m_q=0$ and the {\it total topological charge\/} $Q\ne 0$,
where we define
\disp{Q={1\over 2\pi}\Sigma\bigl[(h^2\curl A_{j+1/2,k+1/2})(\hbox{mod}\ 2
\pi)\bigr].}\noindent
Here the sum is over all plaquettes $(j+1/2,k+1/2)$, and as usual
$X(\hbox{mod}\ 2\pi)=X+2m\pi$, where $m$ is an integer such that $-\pi<
x+2m\pi\le\pi$. Usually (in periodic boundary conditions or vacuum far
field) $Q$ is an integer. According to a special case of the
Atiyah-Singer index theorem (see, e.g., \cite{AtS}) the continuum analog
of Eq.\ (4.2) has $Q$ eigenmodes with zero eigenvalues. Hence, and
because of the doubling effect discussed above, Eq.\ (4.2) will have $2Q$
almost-zero modes (AZMs). In this situation, as explain in Sec.\ 3.9,
these AZMs will not converge in the usual multigrid algorithm, and they
have to be expelled by recombining $2Q+1$ iterants (or only $Q+1$
iterants if one is careful to separately recombine each subsystem).\par
If $m_q>0$ the eigenvalues are shifted away from 0, and then
recombinations need not be done on fine grids, only on coarse ones where
(3.26) still fails, hence their extra computational cost will usually be
small. (If the coarse grid on which recombination is needed is to be
visited many times, it may be more efficient to combine iterants so as to
explicitly isolate and store the AZMs, and then use the latter directly
at each new visit to the grid.)\par
In case a large topological charge is concentrated at few special
plaquettes, it may be ``invisible" to the next coarser grid. This is a
case of a {\it small-scale essential singularity\/}, which may be another
reason for recombining iterations (see Sec.\ 3.8), even if the total $Q$
vanishes.\par
On grids where $h$ approaches $\xi_{_G}$, the algorithm of Sec.\ 4.3
should radically be changed (see Sec.\ 4.5). However, if $\xi_{_G}$ is
not too small compared  with $\min(\xi_m,hL)$, then one can simply afford
solving on such grids by a slower, usual iterative method, such as
Kaczmarz relaxation with conjugate gradient acceleration. In particular,
if such a grid with such a slower solver is taken as the coarsest level
for a multigrid cycle that has several finer levels, then the cycle
efficiency will not be substantially disturbed by it.\par
Indeed, experiment showed that for $m_q$ and $\beta$ in the physically
interesting ranges the standard multigrid efficiency is obtained by this
algorithm (enforced on some levels as in Sec.\ 4.5 below) even without
recombinations \cite{A2}, \cite{PT2}, \cite{PhDR}. More recent
experiments showed that smaller $m_q$ can be accommodated as efficiently
by recombining iterants.\par
Incidentally, the experiments showed that a special care should be taken
in the statistical process that generates the gauge field. A slow
Monte-Carlo process with a cold $(A=0)$ start may never yield $Q\ne 0$,
and with a hot start (random $A$) may get stuck with too large $Q$ and
too short $\xi_{_G}$. This topic belongs of course to Sec.~5 below.\par
\subsection{Scales\/ $h\sim\xi_{_G}\ll\min(hL,\xi_m)$}
The multigrid algorithm described above starts to have troubles when it
is employed for grids whose meshsize $h$ approaches $\xi_{_G}$. Similar
to the case of disordered diffusion problems (Sec.\ 3.5) the main
difficulty has to do with the representation $A^{2h}$ of the gauge field
on the coarser grids.\par
Indeed, the values of $h$ for which the algorithm still works has been
pushed a significant factor up by replacing the naive injection (see
Sec.\ 4.3) with a {\it parallel transport of the gauge field\/} itself,
carefully separating different ``kinds" of links. One ``kind" of links,
for example, connects $\psi^1$ points to neighboring $\psi^3$ points;
another connects $\psi^3$ to $\psi^2$; etc. A link on grid $2h$ should be
composed from a pair of grid-$h$ links {\it of the same kind\/}. Since
the location of such a pair does not necessarily coincides with that of
the grid-$2h$ link, the pair should sometimes be parallel transported
from another location. Such constructions are simpler to formulate if the
coarsening is done one dimension at a time. See \cite{A2} and
\cite{PhDR} for details.\par
\subsection{Disorder\/ at $\xi_{_G}\le h\ll\xi_m$}
As the meshsize becomes larger than $\xi_{_G}$, the system of equations
is stuck in ``disorder" (which could be avoided, though: see Sec.\ 4.7).
Similar to the situation described in \S3.5, the large-scale connections
seem to no longer follow a well-ordered pattern similar to that on the
fine grid. It is thus natural to seek an AMG-like approach in formulating
the grid-$2h$ equations. (Such an attempt has been initiated by R.\
Ben-Av.)\par
In the AMG approach $L^{2h}$ (written out as a {\it real\/} system) is
constructed by the Galerkin form (3.9), where $I_h^{2h}$, in cases of
self-adjoint systems, is chosen by (3.10). This reduces the problem of
coarsening to that of constructing only a good interpolation scheme.\par
However, our system (4.2) is {\it not\/} self-adjoint. It is tempting to
turn it into self-adjoint by replacing (4.2) with
\disp{L^{h\dag}L^h\Psi=L^{h\dag}f.}\noindent
This, however, would raise the order of derivatives in the system from 1
to 2, hence would require, by (3.17), the construction of a higher order
interpolation, which is considerably more difficult. It may be better to
stay with the original system and construct $I_h^{2h}=2^{-d}(\widehat{
I}_{2h}^h)^T$, where $\widehat{I}_{2h}^h$ does not necessarily coincide
with $I_{2h}^h$: in the same way that $I_{2h}^h$ is constructed as a
``good interpolation" (see below) for $L^h$, $\widehat{I}_{2h}$ should be
constructed as a good interpolation for the {\it adjoint\/} of $L^h$.\par
A good interpolation in the AMG approach means not just good
interpolation {\it weights\/}, but also a good choice of the coarse-grid
variables. For both purposes one has to distil local relations which must
be satisfied (to accuracy orders spelled out in (3.17)) by {\it all\/}
error functions that converge slowly under the employed relaxation
scheme.\par
One general approach for finding such local relations has been called
{\it pre-relaxation\/} (see Sec.\ 6.1 of \cite{AMGS}). Applying several
sweeps of the given relaxation scheme to the {\it homogeneous\/} (zero
right-hand side) equations will result in a typical shape of a
slow-to-converge error. (For the homogeneous equations the solution
vanishes, hence the current approximation equals the current error.)
Repeating this for several random initial approximations yields several
such typical error functions, independent on each other even locally. At
each point, a suitable local relation is any relation approximately
satisfied at that point by all these error functions. It can be distilled
from them by usual data fitting (least square) techniques.\par
In this way $L^{2h}$ is constructed from $L^h$. Similarly $L^{4h}$ can
then be constructed from $L^{2h}$, and so on.\par
The same approach could also be used at finer levels, where $h\ll
\xi_{_G}$, but it is much more expensive and less efficient there than
the parallel-transport algorithm described above.
\subsection{Introducing order}
The algebraic multigrid (AMG) approach is very expensive not only in
deriving the coarse level operators (and re-deriving them upon each
change in the gauge field), but also in operating them, especially since
they loose much of the sparsity associated with distinguishing between
different species. Avoiding the disorder that motivates AMG by observing
an underlying apriori order can make the algorithm simpler and much more
efficient. Possible approaches are outlined below.\par
First we observe that gauge fixing, and especially {\it re\/}-fixing, has
negligible cost compared with the Dirac solver itself (cf.\ Sec.\ 4.2).
With the gauge field satisfying (4.7), and hence smooth for $h\ll
\xi_{_G}$, and with species labelled as in Sec.\ 4.3, Eq.\ (4.2) in the
limit $h\rightarrow 0$ gives
\def\theequation{4.10a}
\equa{D_1\Psi^1+D_2\Psi^2+m_q\Psi^4=f^4}
\def\theequation{4.10b}
\equa{-D_2\Psi^1+D_1\Psi^2+m_q\Psi^3=f^3}
and two similar equations with $f^1$ and $f^2$, where
\def\theequation{4.10c}
\equa{D_{\mu}=\ptl_{\mu}-iA^{\mu}.}
Since $h\ll\xi_m$ at all levels for which we need to construct a
multigrid solver (see Sec.\ 4.4), we can assume in describing any
underlying local order that $m_q=0$. Our system breaks then down into two
subsystems decoupled from each other (see Sec.\ 4.3), one of them being
(4.10). Defining
\def\theequation{4.11}
\equa{\Psi^+=\Psi^1-i\Psi^2,\qquad\Psi^-=\Psi^2-i\Psi^1,}
it is easy to see that for $m_q=0$ Eq.\ (4.10) yields
\def\theequation{4.12}
\equa{(D_1\pm iD_2)\Psi^{\pm}=f^{\pm},}
where $f^+=f^4-if^3$ and $f^-=f^3-if^4$. Define further $\Xi^+$ and
$\Xi^-$ to be solutions of
\def\theequation{4.13}
\equa{(\ptl_1\pm i\ptl_2)\Xi^{\pm}=i(A^1\pm iA^2).}
Finally, defining $\Phi^+$ and $\Phi^-$ by
\def\theequation{4.14}
\equa{\Psi^{\pm}=e^{\Xi^{\pm}}\Phi^{\pm}}
and substituting into (4.12), we obtain, by (4.13),
\def\theequation{4.15}
\equa{(\ptl_1\pm i\ptl_2)\Phi^{\pm}=e^{-\Xi^{\pm}}f^{\pm}.}
\par\vskip -\parskip
These relations show the underlying regularity in Eq.\ (4.2), because
both (4.13) and (4.15) are regular elliptic systems, each in fact
equivalent (when written as a {\it real\/} system) to the Cauchy-Riemann
equations (3.19). Multiplying Eq.\ (4.13) by $(\ptl_1\mp i\ptl_2)$, it
can also be written as the Poisson equations
\def\theequation{4.16}
\equa{\Delta\Xi^{\pm}=\pm\curl A+i\hbox{div} A.}
This suggests that for the underlying functions $\Xi^{\pm}$ to be well
approximated on any grid $2h$, the gauge field $A^{2h}$ should be
generated from $A^h$ by requiring $\curl A^{2h}$ and $\hbox{div} A^{2h}$
to be local averages of $\curl A^h$ and $\hbox{div} A^h$, respectively,
and solving for $A^{2h}$ via the algorithm of Sec.\ 4.2. (In particular,
if $\hbox{div} A^h=0$ on the finest grid by gauge fixing, it will remain
so on all coarser grids, and a cut-torus gauge $A^h$ will give cut-torus
gauge fields on all coarser grids, with the same cut and the same jump
$C_{\ast}$.) More important, when $h\ge O(\xi_{_G})$, values of the
topological charge $\vert h^2\curl A\vert$ obtained from finer levels by
such averaging may exceed $\pi$. Such values cannot actually affect
$\Xi^{\pm}$ because by (4.3) $hA$ is only defined modulo $2\pi$. This may
well explain the difficulty encountered at such scales.\par
This also suggests two possible approaches around the difficulty. One is
to treat large concentrated topological charges that cannot be
represented on coarser grids by general methods developed for small-scale
essential singularities (see Sec.\ 3.8). Namely, smear the topological
charge on a wider area at the coarser levels and/or recombine iterants.
\par
Another, more radical approach, which can also treat difficulties arising
from the {\it imaginary\/} part of (4.16), is to abandon (4.2), at least
on coarser levels, and, with the gauge field fixed to satisfy (4.7),
discretize (4.10) directly. In particular, the covariant derivatives
$D_{\mu}^h$ will be obtained by direct central differencing of (4.10c),
circumventing the limitation on the size of $\vert h^2\curl A\vert$.\par
This approach represents the {\it fixed\/} field $A^{\mu}$ as a field
that has a continuous continuum limit, which is generally true only
piecewise. It generally contradicts, for example, periodic boundary
conditions for the field $U^h=e^{ihA}$. This difficulty does not seem to
be substantial; first, because periodic boundary conditions do not
represent any physical reality. Also, for periodic boundary conditions
the above approach can be used {\it patchwise\/}, employing the cut-torus
gauge. Namely, each multigrid process (relaxation, inter-grid transfer)
at any neighborhood can be done with the cut line shifted away from that
neighborhood.\par
In higher dimensions and for non-Abelian gauge fields, the ``gauge phases
per unit length" with the cut-torus fixing still have piecewise
continuous limit, so the described approach seems still applicable.\par
\subsection{Superfast updates. Localization by distributive changes}
In addition to the fast solver, the multigrid structure can yield very
fast procedures for updating the solution upon any local change in the
data.\par
For simplicity we will assume in the discussion here that $m_q=0$; for
larger $m_q$ the range of influence of changes will be shorter, hence the
assertions below will hold even more strongly. Also for simplicity we
will consider the system (4.10), with the gauge field satisfying (4.7);
gauge transformation will not change our conclusions either.\par
Consider first the case of a change introduce to $f_{0,0}^4$, the value
of the forcing term $f^4$ at the origin. In vacuum $(A\equiv 0)$, this
will change the solution $\Psi_{j,k}^h$, at distance $r=h(j^2+k^2)^{1/2}$
from the origin, by an amount which is $O(r^{-1})$. Moreover, the
$\ell$-th derivative (or difference quotient) of the change will decay
like $O(r^{-1-\ell})$. Thus the change is not necessarily very small, but
becomes very {\it smooth\/} at a short distance (few meshsizes in fact)
from the origin. This implies that in applying the FMG algorithm (cf.\
Fig.~2) to the {\it change\/} in the solution, on every grid one has to
employ relaxation only in the neighborhood of the origin (upto few
meshsizes away).\par
Furthermore, one can cut the work far more by introducing the changes,
say in $f^3$, in a {\it distributive\/} manner. This means that changes
are distributed to {\it several\/} values of $f^3$ at a time, according
to a prescribed pattern. For example, changing simultaneously $f_{j,k}^3$
and $f_{j,k-2}^3$ by $+\delta$ and $-\delta$ respectively is a {\it
first-order distributive change\/}. Changing $(f_{j,k-2}^3,f_{j,k}^3,f_{j
,k+2}^3)$ by $(+\delta,-2\delta,+\delta)$ is {\it second-order\/}, and so
is changing $(f_{j-2,k-2}^3,f_{j-2,k}^3,f_{j,k-2}^3,f_{j,k}^3)$ by
$(+\delta,-\delta,-\delta,+\delta)$. Etc. The effect of an $m$-th order
distributive change on the solution will decay as $O(r^{-1-m})$. Hence
with an appropriate choice of distribution order ($m=1$ may indeed
suffice), the effect of the solution becomes practically local, and can
be obtained by few relaxation steps in some neighborhood of the change.
Only once in several such changes on grid $h$ one need to go {\it
locally\/} to the coarser grid (cf.\ Sec.\ 4.2); and once in several such
transfers to grid $2h$, a similar transfer (local on scale $2h$) is made
from grid $2h$ to grid $4h$; and so on. The work per change is just
$O(1)$.\par
The same must be true in a non-vacuous gauge field, as can be seen from
(4.15), except that on grids with meshsize $h\gsim\xi_{_G}$ the functions
$\Xi^{\pm}$ have random second ``derivatives", so the effect of second or
higher order distributive changes is more complicated and requires a
probabilistic investigation.\par
Distributive changes do not span all the changes of interest, but all
is needed to complement them are {\it smooth\/} (on scale $h$) changes.
The latter can be processed on the grid $2h$. Moreover, they too can be
distributive (on scale $2h$), complemented by smooth (on scale $2h$)
changes; etc.\par
Changes in the gauge field itself can be treated similarly: their effect,
as can be seen from (4.16), can also be localized by distribution. If the
changes are to be governed stochastically, one {\it submits\/}
distributive changes to the Monte-Carlo process. This should be
complemented by a corrective Monte-Carlo process on grid $2h$ (as one
needs to do anyway to avoid slowing down of the simulation: see Sec.~5).
The latter can be again distributive, complemented by a grid-$4h$
process; and so on. On each scale such distributive changes have only
local effects, easily established by local relaxation.\par
\subsection{Inverse matrix and determinant}
The multigrid structure can also provide efficient ways for storing,
updating and using information related to the inverse matrix $M^{-1}=(L^h
)^{-1}$. For a large lattice with $n$ sites, the storage of the inverse
matrix would require $O(n^2)$ memory and $O(n^2)$ calculations, even with
a fully efficient multigrid solver. Both can be reduced to $O\bigl((\ell+
\epsi^{-1/\ell})^dn\bigr)$, where $\epsi$ is the relative error allowed
in the calculations and $\ell$ is the interpolation order below, by using
the following multilevel structure.\par
Denoting the propagator from gridpoint $x=(jh,kh)$ to gridpoint
$y=(j^{\prime}h,k^{\prime}h)$ by
\disp{M^{-1}(x,y)=\bigl((L^h)^{-1}\bigr)_{(j,k),(j^{\prime},k^{\prime}
)},}\noindent
the $\ell$-th ``derivatives" (difference quotients) of this propagator,
with respect to either $x$ or $y$, decay as $O(\vert x-y\vert^{-1-\ell}
)$. Therefore, an $\ell$-order interpolation of the propagator from grid
$2h$ to grid $h$ will have at most $O\bigl(h^{\ell}(\vert x-y\vert-\ell
h/2)^{-\ell}\bigr)$ relative error, which will be smaller than $\epsi$ in
the region
\disp{\vert x-y\vert/h\ge K\eqd C\epsi^{-1/\ell}+\ell/2,}\noindent
where $C$ is a (small) constant. Hence, propagators $M^{-1}(x,y)$ with
$\vert x-y\vert\ge Kh$ need be stored on grid $2h$ only, except that, for
a similar reason, those of them with $\vert x-y\vert\ge 2Kh$ need
actually be stored only on grid $4h$; and so on.\par
This structure can be immediately updated, upon changes in the gauge
field, especially if those are made in the above distributive manner
(cf.\ Sec.\ 4.8). Changes is propagators described on grid $2h$
(associated with relaxing the {\it smooth\/} changes in the gauge field)
affect those described on grid $h$ through a FAS-like interpolation (cf.\
Sec.\ 3.3: it means correcting $u^h$ by $I_{2h}^h(u^{2h}-\ov{I}_h^{2h}u^h
)$; except that here one interpolates both in $x$ and in $y$). The cost
per update is $O(1)$, i.e., independent of lattice size.\par
With $M^{-1}$ thus monitored, one can inexpensively calculate changes in
$\log\det M$. For a {\it small\/} change $\delta M$ in the gauge field
\def\theequation{4.17}
\equa{\delta\log\det M=\tr(M^{-1}\delta M),}
which can be computed locally, based on $M^{-1}(x,y)$ with neighboring
$x$ and $y$. For larger changes one can {\it locally integrate\/} (4.17),
since the local processing also gives the dependence of $M^{-1}$ on
$\delta M$. Again, the amount of calculations per update does not depend
on the lattice size.\par
\section{Multiscale Statistical Simulations}
The problem of minimizing a particle energy $E(r)$, or an energy
$E^{\alpha}(u^{\alpha})$ of a function $u^{\alpha}$ defined on a
$d$-dimensional lattice with meshsize $\alpha$, has been used in Sec.~1
to introduce several multiscale processes. Similar processes can also be
very useful in accelerating statistical simulations governed by such an
energy (or Hamiltonian) $E^{\alpha}$. The first objective of such
simulations is to produce a random sequence of effectively independent
configurations $u^{\alpha}$, in {\it equilibrium\/}, i.e., in such a way
that the probability of any $u^{\alpha}$ to appear anywhere in the
sequence is given by the Boltzmann distribution
\def\theequation{5.1}
\equa{P(u^{\alpha})={1\over Z}e^{-E^{\alpha}(u^{\alpha})/T},}
where $T$ is the temperature and $Z$ is a normalization factor such that
$\int P(u^{\alpha})Du^{\alpha}=1$.\par
The minimization problem is in fact easily seen to be the limit
$T\rightarrow 0$ of (5.1). It has been shown in Sec.~1 that multiscale
processes can solve this limit problem in $O(n)$ computer operations,
where $n=L^d$ is the number of lattice sites. In a similar way it will be
shown here that the multiscale-accelerated simulation needs only $O(n)$
operations to produce each new, effectively independent configuration.
\par
Such accelerations were first introduce, independently, in \cite{AS3} and
in Sec.\ 7.1 of \cite{A18}. (A significant difference is that in
\cite{AS3} the ``constant interpolation" is used. The implications of
this will be examined below.) Another type of acceleration was introduced
for Ising spin models in \cite{SW}, and then extended by embedding
\cite{AS55} to many other models (see review in \cite{AS31}). The
relation between, and combination of, these two types of acceleration
will be outlined below.\par
The central claim, however, of this chapter (following \cite{A5}) will
be that accelerating the production of effectively independent
configurations is not exactly the main issue. The real objective of the
statistical simulations is to calculate some average properties of the
configurations $u^{\alpha}$, and the main issue is how fast deviations
from any desired average can be averaged out. The multigrid structure
will be shown very useful in cheaply providing much statistical sampling
in its coarse levels, thus promoting fast averaging of large scale
fluctuations, which are exactly the kind of fluctuations not effectively
self-averaged in any one produced configuration.\par
In usual Monte-Carlo processes, one useful measurement cost $O(L^{d+z})$
operations, where typically $z\approx 2$. The acceleration techniques
ideally remove the Critical Slowing Down (CSD), i.e., the factor $L^z$.
For the ideal Gaussian case it has been shown \cite{A13} that using
coarse-level sampling can eliminate the volume factor $L^d$ as well. The
development of such techniques to other models, including spin models, is
discussed below.\par
\subsection{Multiscale Monte-Carlo: unigrid}
Similar to the point-by-point relaxation in Sect.\ 1.5, the usual way to
simulate (5.1) is the point-by-point {\it Monte-Carlo\/} process. The
basic step is to simulate one variable $u_i^{\alpha}$: holding all other
$u_j^{\alpha}$ fixed, (5.1) describes a probability distribution for
$u_i^{\alpha}$, which can easily be simulated, e.g., by assigning to
$u^{\alpha}$ a random value with that distribution. A {\it point-by-point
Monte-Carlo sweep\/} is the repetition of the basic step at all sites
$(i=1,2,\ldots,n)$. A long enough sequence of such sweeps will produce a
new (effectively independent) configuration, the probability distribution
of which is indeed (5.1).\par
And similar to the point-by-point relaxation, the main trouble of this
Monte-Carlo process is its slowness: typically, for a lattice with
$n=L^d$ sites, a sequence of $O(L^2)$ {\it sweeps\/} is needed to produce
a new configuration. The same smooth components which are slow to
converge in any local relaxation, are also slow to change in any local
Monte-Carlo process, and for similar reasons. So steps of more collective
nature are required here as well.\par
{\it A Monte-Carlo step on scale\/} $h$ is a collective move of the form
(1.12), whose amplitude $u_k$ is decided stochastically, under the
probability distribution deduced for it from (5.1). {\it A Monte-Carlo
sweep on scale\/} $h$ is the repetition of such a step at all points
$x_k$ of a grid with meshsize $h$ $(h>\alpha)$, laid over the original
grid $\alpha$. {\it A\/} ({\it unigrid\/}) {\it multiscale Monte-Carlo
cycle\/} is a process that typically includes a couple of Monte-Carlo
sweeps on each of the scales $\alpha,2\alpha,\ldots,2^{\ell}\alpha$,
where a sweep on scale $\alpha$ is just the usual point-by-point
Monte-Carlo sweep, and $2^{\ell}\alpha$ is a meshsize comparable to the
linear size of the entire lattice. Under ideal situations, each such
cycle will produce a nearly independent configuration.\par
Such cycles were introduced in \cite{HM5}, \cite{HM8} under the name
``multigrid". We will use for them the adjective ``unigrid" to emphasize
that all the moves are still performed in terms of the finest grid (the
given lattice), thus distinguishing this process from the more developed
multigrid; cf.\ the comment at the end of Sec.\ 1.3.\par
Compared with the more developed multigrid that will be described later,
the unigrid cycle has two basic disadvantages. First, the moves on coarse
scales are very expensive: each move (1.12) on scale $h$ involves
changing $O\bigl((h/\alpha)^d\bigr)$ values in the basic grid. This
disadvantage is not so severe for cycles that employ the same number of
sweeps at all scales, since the number of moves in each sweep on scale
$h$ is just an $O\bigl((h/\alpha)^{-d}\bigr)$ fraction of their number in
each sweep on the basic grid. But we will see below that for statistical
purposes one would better do many more sweeps on coarse scales than on
fine ones, so this disadvantage will become crucial.\par
A second, not less serious disadvantage is that it is often impossible to
prescribe in advance the shape functions $w_k^h(\zeta)$ that control the
large-scale moves (1.12) so that high enough probabilities will result
for producing reasonably large amplitudes $u_k^h$. Suitable shape
functions can be prescribed apriori only if the shapes of the probable
large scale moves are indeed sufficiently independent of the current
configuration (see also Sec.\ 5.6).\par
Even when such apriori probable shape functions exist, their calculation
is often best obtained in the multigrid manner, in which each level of
shape functions $w_k^h$ is derived from the next-finer-level shapes
$w_k^{h/2}$. We have seen the importance of such derivation even for {\it
deterministic\/} problems, e.g., in Secs.\ 3.5 and 4.6 above, where the
interpolation $I^h_{2h}$ between neighboring scales is non-trivial and
need be separately derived at each level. For {\it stochastic\/} problems
the need for such a hierarchical construction of large-scale moves is
much stronger, because of the statistical dependence between moves at
different scales, especially neighboring scales.\par
On the other hand, the unigrid approach has the important advantage that
it does not require derivations of the coarse level Hamiltonians, which
can be quite problematic: see Sec.\ 5.5.\par
\subsection{Multigrid Monte-Carlo}
Thus, instead of performing the moves directly in terms of the finest
grid, the multigrid approach, similar to Sec.\ 1.6 above, is to consider
the moves $u^{2h}$ on any grid $2h$ as a {\it field\/} which {\it
jointly\/} describes displacements for the next finer field, $u^h$, by
the relation
\def\theequation{5.2}
\equa{\delta u^h=I_{2h}^hu^{2h},}
where $I_{2h}^h$ is, as before, an operator of interpolation from grid
$2h$ to grid $h$. Assume for now that the grid-$2h$ Hamiltonian, at any
given fine-grid configuration $u^h$,
\def\theequation{5.3}
\equa{E^{2h}(u^{2h})=E^h(u^h+I_{2h}^hu^{2h})}
has been derived as an {\it explicit\/} function of $u^{2h}$ (with
coefficients possibly depending on $u^h$). Then the multigrid cycles
described above (Fig.~1) can be employed here; the only difference being
that ``relaxation sweeps" are replaced by ``Monte-Carlo sweeps".\par
Under ideal situations, each cycle, with $\nu_1+\nu_2=2$ or 3, would
produce a new, effectively independent configuration. More precisely we
mean by this that the correlation between any quantity of interest in the
initial configuration and in the one produced after $k$ cycles decays
like $e^{-k/\tau}$, where $\tau$, the {\it cycle auto-correlation
time\/}, is independent of the lattice size $L$; indeed $\tau$ is often
smaller than 1. Observe that as long as the cycle index $\gamma$ is less
than $2^d$, most of the work in each cycle is just the $\nu_1+\nu_2$
Monte-Carlo sweeps on the finest lattice. So the total work for
producing a new configuration is just $O(L^d)$, or just proportional to
the number of sites in the lattice. By comparison, in the {\it
unigrid\/} approach this work is $O\bigl((\log L)L^d\bigr)$ for $\gamma=
1$ and $O(L^{d+\log_2\gamma})$ for $\gamma>1$.\par
\subsection{Gaussian model: Eliminating CSD}
The prime example of the ``ideal situation" is the Gaussian model, with
the Hamiltonian
\def\theequation{5.4}
\equa{E^{\alpha}(u^{\alpha})=\sum_{\langle j,k\rangle} a_{j,k}^{\alpha}
(u_j^{\alpha}-u_k^{\alpha})^2,}
where the summation is over pairs of {\it neighboring\/} sites $j$ and
$k$, and $a_{j,k}^{\alpha}$ are non-negative coupling coefficients. This
Hamiltonian could arise as a discretization of (3.2). Since the
interpolation $I_{2h}^h$ is a linear operator, coarse grid Hamiltonians
(5.3) can easily be derived and will have again the form (5.4), except
that the range of neighbors $k$ (such that $a_{j,k}^{2h}\ne 0$) for each
site $j$ will depend on the interpolation order. In the particular
Gaussian case the coefficients $a_{j,k}^{2h}$ of the level-$2h$
Hamiltonian $E^{2h}$ will {\it not\/} depend on the current fine-grid
configuration $u^h$; they will only depend on the next-finer grid {\it
coefficients\/} $a_{j,k}^h$, and on the coefficients of the interpolation
operator $I_{2h}^h$. Thus, for a given $E^{\alpha}$, the coarsening
process depends only on the choice of the interpolation operators, and so
is also the efficiency of the entire multigrid cycle.\par
To obtain the ideal efficiency in this case, the coarser levels should
accurately sample all the components slow to change under the
current-level Monte-Carlo process. This implies that every slowly
changing configuration $v^h$ must have an approximate ``{\it coarse-grid
representation\/}", i.e., an approximate configuration of the form
$I_{2h}^hu^{2h}$, and the two configurations should have approximately
the same energy.\par
In case of {\it smooth and isotropic\/} coefficients (e.g., $a_{j,k}^h$
depending only on the distance from $j$ to $k$), the slow-to-converge
components are simply the {\it smooth\/} components, which can indeed be
approximated by interpolants from a coarser grid. The requirement of
approximating the energy implies in this case that the order of
interpolation should be at least 2, i.e., linear or multi-linear
interpolation, such as given by (1.7). (The required order is in fact a
special case of the rule (3.17) above.) Indeed, with such an
interpolation, a multigrid $V$ cycle (see Fig.~1) is so efficient that it
is hard to measure any correlation between the susceptibilities before
and after the cycle. (The exact value of the very small autocorrelation
time $\tau$ depends on the details of the simulation at the coarsest
level, and is not important anyway.)\par
A border case is the {\it first\/} order {\it constant interpolation\/}
$I_{2h}^h$, defined by the shape function
\def\theequation{5.5}
\equa{\begin{array}{ll}
&w_k^{2h}(\zeta_1,\ldots,\zeta_d)\\
&\qquad =\left\{\begin{array}{ll}
1&\hbox{if all}\ \theta_i-1<\zeta_i\le\theta_i\\
0&\hbox{otherwise}
\end{array}\right.
\end{array}}
with any convenient $\theta_1,\ldots,\theta_d$. For any smooth function
$v^h$ approximated by a coarse function $u^{2h}$, the energy of $I_{2h}^h
u^{2h}$ is about {\it twice\/} that of $v^h$. Hence, a component
$v^{\alpha}$ on the finest grid which is so smooth that it effectively
changed only on a coarse grid with meshsize $2^m\alpha$, say, will be
represented for that coarse grid by an approximation which has an energy
$2^m$ times its own energy. Therefore, that coarse grid will change such
an approximation a change with size only $O(2^{-m/2})$ the probable size
for changes of that smooth component. Hence, roughly $2^m$ visits to that
coarse grid will be needed to (randomly) accumulate the actual probable
size of a change. This means that any grid $2h=2^m\alpha$ should be
visited at least twice per each visit to grid $h=2^{m-1}\alpha$, i.e.,
{\it a cycle index\/ $\gamma\ge 2$ must be used\/}.\par
Experiments \cite{AS3} indeed show that a $W$ cycle $(\gamma=2)$ is
enough to produce $\tau=O(1)$, while a $V$ cycle $(\gamma=1)$ is not. For
dimension $d\ge 2$ this algorithm with constant interpolation still
eliminates CSD, since the work per cycle is still $O(L^d)$ for any
$\gamma<2^d$.\par
When the ``diffusion coefficients" $a_{j,k}^{\alpha}$ are anisotropic or
wildly changing in size, the interpolation operators that produce good
approximations to slowly changing components get more complicated. In
cases of consistent anisotropy, semi-coarsening (e.g., decimation only in
the direction of strong couplings) should be used (cf.\ Sec.\ 4.2.1 in
\cite{A10}). If the coefficients change wildly, shape functions and
coarsening strategies similar to those in Sec.\ 3.5 above need be
employed.\par
It need perhaps be emphasized that the multigrid process eliminates CSD
from Gaussian models with {\it variable\/} coefficients, which {\it
cannot\/} generally be done by Fourier methods. Also, unlike Fourier,
general non-periodic domains can be handled with the same efficiency.\par
Non-Gaussian models will be discussed in Sec.\ 5.5. First, however, we
will argue in the next section that elimination of CSD is not the only,
perhaps not even the most important, issue.\par
\subsection{Eliminating the volume factor}
In usual statistical simulations on a $d$-dimensional grid of size $L^d$,
the amount of computer operations needed to produce one statistically
independent measurement is $O(L^d\xi^z)$, where $\xi$ is the correlation
length, which is normally $O(L)$. The exponent $z$ is called the {\it
dynamic critical exponent\/}. Typically $z\approx 2$ for point-by-point
Monte-Carlo methods. Eliminating the critical slowing down, i.e., the
factor $\xi^z$, has been obtained by multigrid, as discussed above, and
also by other methods. The important advantage of the multigrid approach
is that it can potentially drastically reduce the {\it volume factor\/}
$L^d$ as well. For Gaussian models it has been demonstrated \cite{A13}
that this factor, too, can be {\it completely eliminated\/}. This is
especially good news for high dimensional (e.g., $d=4$) problems.\par
Statistical fluctuations in physical systems occur on different scales:
there are local fluctuations, intermediate-scale fluctuations,
large-scale ones. Generally they are not independent of each other;
especially, fluctuations at neighboring scales can be highly correlated.
But in many cases there is only weak correlations between deviations at
two widely different scales.\par
At the coarse levels $h>2^m\alpha$ of a multigrid cycle, the finer-scale
fluctuations are effectively frozen at their values in the current
configurations $u^{\alpha},\alpha^{2\alpha},\ldots,u^{2^m\alpha}$. The
coarser-scale fluctuations, if dependent only weakly on those frozen, can
be averaged out on the coarse levels alone, before any return to the
finer levels, by letting the coarse level Monte-Carlo simulation be
suitably long and accompanied by a suitable sequence of measurements.
Such averaging out of large scale fluctuations can be very efficient
because on these coarse grids such fluctuations are sampled rapidly
(large changes per sweep) and cheaply (little work per sweep).\par
This of course depends on having a good enough representation of smooth
components on the coarse grids. This for example is {\it not\/} the case
if, in the Gaussian or other asymptotically free models, {\it constant\/}
interpolation is used in the inter-grid transfers. As explained in Sec.\
5.3, such an interpolation represent smooth components by far more
energetic relatives, which therefore can move only a small fraction of
the movements typical to the smooth components, not enough to allow
averaging out of fluctuations. {\it Linear\/} interpolation, on the other
hand, does represent the large-scale fluctuations on the correspondingly
coarse grids by components of nearly the same energy, hence allow their
full averaging by the coarse grid Monte-Carlo.\par
The {\it fine-scale\/} fluctuations are largely {\it self-averaged\/} in
each given fine-grid configuration. That is, local fluctuations at
different parts of the grids are nearly independent, hence provide nearly
independent samples. These samples are rapidly and cheaply changed by the
Monte-Carlo process on that grid.\par
Thus, generally, it takes only $O(1)$ work to replace any sample of a
fluctuation on any given scale by the Monte-Carlo sweeps at the
corresponding meshsize. Hence, if measurements accompany the simulation
closely enough, fluctuations on any scale can be averaged out very
efficiently.\par
The relative number of Monte-Carlo sweeps needed at each scale $h$
depends on how much averaging-out is needed for the fluctuations at that
scale, which is roughly $O(\sigma_h^2)$, where $\sigma_h$ is the average
contribution of such fluctuations, in any one configuration, to the
measurement deviation. This contribution depends on the desired
observable. For some (perhaps less interesting) observables, such as
energy, the contribution of finer scales dominate in such a way that most
of the simulation work should be done on the finest grid. In such cases a
cycle index $\gamma<2^d$ should be used.\par
For many (perhaps the more interesting) observables, such as
magnetization or susceptibility (except at $d\ge 6$), the contribution
$\sigma_h$ increases with the scale $h$ in such a way that most of the
Monte-Carlo work should be done on the coarsest grids. This is obtained
by multigrid cycles with index $\gamma>2^d$. In such cases, the finer the
level the more rare its activation, and the finest grid to be reached at
all depends on the accuracy desired for the observable. This indeed
should be the situation with any observable which has a thermodynamic
limit, and the deviations $\sigma_h$ should then more properly be defined
as deviations from that limit, not from the average on any finite
lattice.\par
All these claims are strictly true for the Gaussian model with linear
interpolation, for which the claims were precisely defined and confirmed,
both by detailed numerical experiments and by mode analyses \cite{A13}.
For example, it has been shown that a multigrid cycles with index $2^d<
\gamma<64$ calculate the thermodynamic limit of the susceptibility to
within accuracy $\epsi$ {\it in\/ $O(\sigma^2/\epsi^2)$ computer
operations\/}, where $\sigma$ is the susceptibility standard deviation.
\par
This efficiency --- obtaining accuracy $\epsi$ in $O(\sigma^2/\epsi^2)$
operations --- is the {\it ideal statistical efficiency\/}. It is just
the same relation of computational cost to accuracy as in calculating by
statistical trials any simple average, such as the frequency of ``heads"
in coin tossing. Obtaining this ideal statistical efficiency in the
calculation of thermodynamic limits should generally be the goal of our
algorithmic development.\par
Note in the example described above that the use of $\gamma>2^d$
contradicts the condition stated in Sec.\ 5.3 for eliminating CSD. This
emphatically shows that the main issue is not the CSD, but the overall
relation of computational cost to obtained accuracy. To be sure, any
multigrid algorithm that can achieve the ideal statistical efficiency
would also be able, upon change of $\gamma$, to eliminate CSD.\par
The size $L^d$ of the finest grid that should be employed increases of
course with the decrease of $\epsi$, because one needs to have a grid for
which the computed average is only distance $\epsi$ from its
infinite-grid value. For some observables the dependence $L=L(\epsi)$ may
be such that $L(\epsi)^d$ increases faster than $\epsi^{-2}$. Even in
such cases the ideal statistical efficiency may still be attained, by the
process of {\it domain replication\/}: On a domain with only $O(\epsi^{-
1})$ sites and with periodic boundary conditions, the finest level with
meshshize $\alpha$ is first equilibrated (fast, by a multilevel process),
then coarsened to meshsize $2\alpha$. Using the periodicity, the
$2\alpha$ lattice is now duplicated in each direction (to an overall
$2^d$ factor increase in volume), together with the Hamiltonian $E^{2
\alpha}$. Having equilibrated this wider and coarser lattice, one then
proceeds to meshsize $4\alpha$, where the domain is duplicated again. And
so on until the size of domain needed for accuracy $\epsi$ is obtain, at
which our regular multigrid cycle, accompanied with measurements at
coarse levels, can be performed. On one hand this domain replication
process receives enough averaging information from the finest levels to
have its coarser level Hamiltonians accurate to within $O(\epsi)$. On the
other hand it employs, on coarse grids, the full size of a domain needed
to produce the observable to an accuracy $\epsi$.\par
\subsection{Non-Gaussian models}
It is not at all clear that the same kind of ideal statistical efficiency
can be obtained for interesting models far from the Gaussian. But a
detailed examination of the looming difficulties indicates that they are
not insurmountable.\par
The first difficulty is in deriving the {\it explicit\/} Hamiltonian
$E^{2h}(u^{2h})$ that will satisfy (5.3). One {\it advantage\/} of the
constant interpolation $I_{2h}^h$ (cf.\ Sec.\ 5.3) is that it more easily
yields such an explicit Hamiltonian. However, for the ideal efficiency,
as explained above, {\it linear\/} interpolation seems necessary. For
linear interpolation the explicit expression of $E^{2h}$, $E^{4h}$, etc.\
will get increasingly complicated, defeating the purpose of $O(1)$
calculation per gridpoint in all coarse-level Monte-Carlo processes. A
method to derive simple but {\it approximate\/} explicit coarse-grid
Hamiltonian $E^{2h}(u^{2h})$ has been described in Sec.\ 1.4, based on
the observation that one is only interested in having a good
approximation for {\it smooth\/} $u^{2h}$, i.e., $u^{2h}$ with small {\it
strains\/} $u_k^{2h}-u_{\ell}^{2h}$, at any neighboring sites $k$ and
$\ell$. In case of {\it gauge fields\/}, the relevant strains have the
form $\curl A^{2h}$ (cf.\ Sec.\ 4.1). The Taylor expansion in terms of
small strains, such as (1.9), gives us also {\it strain limits\/}, i.e.,
bounds on the size of the strains under which the truncated expansion
still yields a certain accuracy $\epsi_t$.\par
In many cases one likes $E^{2h}$ to preserve the topological properties
(e.g., $2\pi$-periodic dependence, as in (4.4)) of $E^h$, so that it can
allow large-scale moves characteristic to the topology. Then the Taylor
expansion, such as (1.19), should be approximated again by (e.g.\
trigonometric) functions carrying this topology. The strain limits would
guarantee that this approximation, too, has $O(\epsi_t)$ accuracy. Often,
the approximate $E^{2h}$ derived this way would have the same functional
form (hence the same programs) as $E^h$, but with coefficients and an
external field that depend on the current fine-grid configuration $u^h$
(which of course remains fixed throughout the simulation on grids $2h$
and coarser).\par
Unlike the situation in Sec.\ 1.4, however, in the statistical context
discussed here, the fact that $E^{2h}$ is only approximated may destroy
the {\it detailed balance\/} of the simulation, putting its statistical
fidelity in question. Detailed balance, though, is not sacred: like
everything else submitted to numerical simulations --- like the
continuum, the unboundedness of domains, or the infinite length of the
Monte-Carlo chain, etc.\ --- it can be approximated, {\it provided one
keeps a handle on the approximation\/}. Such a handle, for example, is
the $\epsi_t$ introduced above: it determines the above mentioned strain
limits, which can be actually {\it imposed\/}, and one can always examine
how a further reduction of $\epsi_t$ affects the calculated average.
Generally, $\epsi_t$ will be lowered together with $\epsi$, the target
accuracy of the calculated average. Another handle on the accuracy of
$E^{2h}$ can be the order of interpolation $I_{2h}^h$.\par
The lowering of $\epsi_t$ together with $\epsi$ can be done without
running into slowing down because a given move on a {\it fixed\/} level
becomes smoother and smoother on the scale of the finest grid as the
latter becomes finer and finer, hence this move will be acceptable for
smaller and smaller $\epsi_t$ as $\epsi$ is reduced further and further.
Other ingredients in avoiding slowing down are the ``updates", the
Hamiltonian dependent interpolation and the stochastic appearance of
disconnections, all described next.\par
The Monte-Carlo process on each level is constrained by the strain
limits, inherited from all the coarsening stages leading from the finest
level to the current one. Hence, if the strain limits are approached too
closely at some points of some intermediate levels, the process on
coarser levels will be completely paralyzed. To avoid this, the algorithm
uses {\it updates\/}. An update is a return from any current level $h$ to
the next finer grid, $h/2$, introducing here the displacements implied by
the current grid solution (step (iv) in Sec.\ 1.6), and then coarsening
again (with displacements $u^h$ and strains being now defined with
respect to the {\it updated\/} fine grid solution $u^{h/2}$). This
updates $E^{2h}$ and ``relieves" it from strains too close to their
limits. An update can be done locally, wherever strain limits are
approached. Or, more conveniently and sometimes more effectively, it can
be done globally, e.g., after each full Monte-Carlo sweep. In principle,
while introducing the displacements on grid $h/2$ during an update, some
of that grid strains may approach {\it their\/} limits, requiring an
update at level $h/4$. This, however, seldom happens and cannot cascade
to ever finer levels, because moves on any level are very smooth on the
scale of much finer levels, and will therefore affect their strains very
little. In some models, once in a (long) while a coarse level may cause a
``break", a discontinuous change that requires updating all the way to
the finest scales, but such updates will hopefully be local and
sufficiently rare.\par
Due to such updates, large moves are permissible on coarser grids. To
make such moves also {\it probable\/}, the interpolation $I_{2h}^h$ at
each level should be so constructed so that its potential displacements
are as probable as possible. This implies, for example, that if a certain
variable $u_i^h$ is coupled by $E^h$ more strongly to its neighbors in
one direction than in another, then the interpolation to site $i$ should
have a proportionately larger weight in that direction (cf.\ Sec.\ 3.5).
It is true that on the finest level $\alpha$ the Hamiltonian is usually
isotropic and hence $I_{2\alpha}^{\alpha}$ can be isotropic too; but on
coarser levels, due to stochastic variations in $u^h$ at each coarsening
stage, the produced Hamiltonian $E^{2h}$ is no longer isotropic, hence
anisotropic interpolation $I_{4h}^{2h}$ should correspondingly be
introduced. This tends to create even stronger anisotropy at
corresponding points of still coarser levels.\par
Thus, stochastically, at points of sufficiently coarse grids, the
interpolation may become heavily one-sided, approaching in fact the
constant interpolation. The latter does not entail strain limits. It also
involves deletion of couplings in some directions. These together free
the coarse levels to perform moves that introduce ``topological" changes
(vortices, instantons, etc.) to the configuration.\par
To obtain the ideal statistical efficiency, another concern associated
with non-Gaussian models is the dependence between fluctuations at
neighboring scales (see Sec.\ 5.4). The above blueprint does have the
potential of dealing with that, because the operation at each level does
involve frequent updates from the next finer level(s). In fact, the
assertion made above that an update cannot cascade unboundedly to ever
finer levels is exactly related to the assumption of weak dependence
between far scales.\par
\subsection{Stochastic coarsening. Discrete models}
Note in the above outline that $I_{2h}^h$ never depends on the current
configuration $u^h$; it only depends on the coefficients of $E^h$. This
is necessary for approaching detailed balance as $\epsi_t\rightarrow 0$.
On the other hand, the coefficients of $E^h$ do depend on the current
configuration $u^{h/2}$. Thus $I_{2h}^h$, and hence $E^{2h}$, will also
depend on $u^{h/2}$. This enables the shape of the large-scale moves to
develop stochastically. Far from the Gaussian, such stochastic
development of the collective modes, having them chosen by the system
itself, is essential for making them associated with enough free energy,
hence probable. (This, incidentally, is exactly the important capability
lacking in the unigrid approach; cf.\ Sec.\ 5.1.) By contrast, on the
one hand, an {\it apriori\/} large scale movement, inconsiderate of the
current fine scale fluctuations, is likely to contradict them at many
spots and hence to increase the energy very much; it will thus be
rejected by the Monte-Carlo process (or accepted with a very small,
useless amplitude). On the other hand, large-scale moves which {\it
are\/} directly based on the current configuration, will destroy
statistical fidelity.\par
This is most visible in discrete-state models, such as the Ising and,
more generally, the Potts spin models, which are as far as one can get
from the Gaussian. Consider for example the ferromagnetic Ising model
Hamiltonian
\def\theequation{5.6}
\equa{E(s)=-\sum_{\langle i,j\rangle} J_{ij}s_is_j,\qquad (J_{ij}>0)}
where $s_i=\pm 1$ is the spin at site $i$ of a $d$-dimensional lattice
and the summation is over neighboring $i$ and $j$. At some interesting
(e.g., the critical) temperature $T$, any probable configuration would
exhibit large regions of aligned spins. The only type of a large-scale
Monte-Carlo step for this model is offering the flipping of a large block
of spins, to be accepted or rejected according to the probability
distribution (5.1). If the block is chosen {\it apriori\/}, independently
of the current configuration, its boundary is unlikely to have much
intersection with the current boundaries of regions of aligned spins.
Therefore, flipping the block would most likely add many {\it violated
bonds\/} (negative $J_{ij}s_is_j$), which would increase the energy by
much, hence will most probably be rejected. On the other hand, choosing a
block with boundaries coinciding with the current boundaries of spin
alignment would create statistical bias, favoring moves that increase
magnetization.\par
The solution to this dilemma is indeed stochastic coarsening. An example,
now classical, is the Swendsen-Wang (SW) coarsening \cite{SW}. This
consists of a step-by-step blocking. At each step one positive bond $J_{i
j}$ is ``{\it terminated\/}", i.e., replaced by either 0 ({\it deleted\/}
bond) or $\infty$ ({\it frozen\/} bond, blocking $s_i$ and $s_j$
together) in probabilities $P_{ij}$ and $1-P_{ij}$ respectively. In case
of freezing, $s_i$ and $s_j$ effectively become one spin, whose bonds to
neighboring spins can easily be calculated, yielding a new Hamiltonian,
still having the general form (5.6). It can be proved that if $P_{ij}=
q_{ij}\exp(-J_{ij}\wt{s}_i\wt{s}_j/T)$, where $\wt{s}$ is the current
(termination-time) configuration and $q_{ij}$ does not depend on
$\wt{s}$, then simulating thereafter with the new Hamiltonian preserves
the overall statistical equilibrium. At the next step a positive bond of
the (new) Hamiltonian is similarly terminated, and so on. If $q_{ij}=\exp
(-J_{ij}/T)$, then only spins currently having the same sign will be
blocked together, so that current boundaries of spin alignment will
eventually become part of block boundaries, overcoming the above dilemma.
\par
In the original and most used version of the SW algorithm, the blocking
steps continue until a Hamiltonian with no positive bond is reached. Each
block of spins is now flipped in probability $1/2$. This may be followed
by several usual Monte-Carlo sweeps with the original Hamiltonian (5.6),
and then a new, similar sequence of stochastic blocking steps is made,
starting from the original Hamiltonian. And so on. This algorithm proved
very efficient: the dynamic critical exponent $z$ has been drastically
reduced, although not quite to $z=0$. (Recent measurements \cite{AS74}
indicate it is even lower than the original estimate $z=.35$.)\par
Note that the stochasting blocking is not unlike the constant
interpolation which happens to stochastically develop at coarse levels
in the procedures described in Sec.\ 5.5. Indeed, {\it explicit\/}
stochastic steps in choosing the interpolation $I_{2h}^h$ can be added to
those procedures. This may prove necessary wherever (e.g., at some coarse
levels) the possible local shapes of slowly-changing components belong to
several nearly-disjoint sectors, implying several discrete alternatives
in constructing $I_{2h}^h$.\par
The SW algorithm, and its single cluster variant \cite{AS55}, have been
cleverly generalized to many more models by embedding Ising variables in
those models \cite{AS55}, \cite{AS58}, \cite{AS59}; see review in
\cite{AS31}. These procedures are not built hierarchically as the
multiscale and multigrid algorithms described in our earlier sections,
but often achieve comparable efficiency in reducing $z$. The explanation
is that, due to the discreteness of the Ising variables, blocks are
created of all sizes, thus producing moves on all scales. On the other
hand, a more deliberate multigrid-like organization may produce two
additional benefits. First, it can make the algorithm even more efficient
in reducing the auto-correlation time $\tau$. Secondly, and more
important, it may allow cheap collection of many measurements at the
coarse levels, possibly reducing or eliminating the volume factor as well
(cf.\ Sec.\ 5.4). We will now examine this possibility.\par
\subsection{Multiscale blocking}
The SW algorithm described above can be modified into a hierarchical
multiscale procedure in the following way. At the first level of
coarsening, only a subset of bonds is terminated. This subset is chosen
adaptively so that only blocks of size not greater than $b$ are created
(e.g., $b=2$ or $b=2^d$). The resulting Hamiltonian $E^1$ still includes
many positive bonds (``live interactions"). In the second level of
coarsening, additional bonds are similarly terminated, yielding a
Hamiltonian $E^2$. Etc. The Hamiltonian produced at the $\ell$-th level
still has the form
\def\theequation{5.7}
\equa{E^{\ell}(s^{\ell})=-\sum J_{ij}^{\ell}s_i^{\ell}s_j^{\ell},}
but each ``level-$\ell$ spin" $s_k^{\ell}=\pm 1$ is actually a block of
between 1 and $b$ level-$(\ell-1)$ spins $s_m^{\ell-1}$ having the same
sign (which will be taken as the sign of $s_k^{\ell}$ as well). The
coarsest level Hamiltonian is reached when no positive bond is left.\par
With these levels, the usual multigrid cycles can be applied; e.g., the
cycles displayed in Fig.~1, re-interpreted as follows. Circles at level
$2^{\ell}\alpha$ stand for Monte-Carlo sweeps with the Hamiltonian
$E^{\ell}$. The downward arrow (the coarsening step) from level $2^{\ell}
\alpha$ to $2^{\ell+1}\alpha$ represents the blocking steps that create
$E^{\ell+1}$. The upward arrow (uncoarsening) from level $2^{\ell+1}
\alpha$ to $2^{\ell}\alpha$ means executing in terms of $s^{\ell-1}$ the
flips found in $s^{\ell}$.\par
It is easy to see that a $V$ cycle $(\gamma=1)$ with no Monte-Carlo
passes at coarse levels is exactly equivalent to the original SW
algorithm. By choosing $b>2$, a $W$ cycle $(\gamma=2)$ will not be much
more expensive, and experiments show that it significantly cut the
auto-correlation time $\tau$ \cite{Kan1}, \cite{Kan2}. The impression
indeed was that such a $W$ cycle completely eliminated CSD, yielding
$z=0$, but it has later been proved \cite{AS13} that at least a certain
version of this algorithm (where the bonds terminated first are the same
at all cycles) still suffers a (very marginal) slowing down.\par
However, as emphasized in Sec.\ 5.4, CSD is not the main issue. The
question is how much computational work is needed per
effectively-independent {\it measurement\/}. Consider for example
measurements for the susceptibility $\langle M^2\rangle$, which can be
taken at any level $\ell$ since $\Sigma_k s_k^{\ell}=\Sigma_m s_m^{\ell-
1}=\cdots=\Sigma_i s_i=M$. Can one benefit from averaging over $M^2$ {\it
within\/} the cycle?\par
To sharpen this question, let us denote by $\chi_0=\langle M^2\rangle$
the true susceptibility, by $\sigma_0=\langle(M^2-\chi_0)^2\rangle^{1/2}$
the standard deviation from $\chi_0$ of $M^2$ at any single
configuration, by $\chi_1$ the average of $M^2$ for the Hamiltonian $E^1$
and by $\sigma_1=\langle(\chi_1-\chi_0)^2\rangle^{1/2}$ the standard
deviation of $\chi_1$ from $\chi_0$. The question then boils down to
this: is $\sigma_1$ much smaller than $\sigma_0$? Does $\sigma_1/
\sigma_0\rightarrow 0$ as $L\rightarrow\infty$?\par
The first answer to this question was disappointing. Two dimensional
experiments near the critical temperature showed that as $L$ increases,
both $\sigma_1$ and $\sigma_0$ remain proportional to $\chi_0$; the ratio
$\sigma_1/\sigma_0$ is determined only by the {\it fraction\/} of bonds
being terminated in creating $E^1$. In reducing the number of degrees of
freedom one looses not just fine-scale fluctuations, but large-scale ones
as well.\par
At first, this strong correlation between scales appears to be a
necessary property of discrete-state models. But then, a similar
situation is encountered when {\it constant\/} interpolation is used even
for the Gaussian model (cf.\ Sec.\ 5.3). So the question now is whether a
better coarsening technique, capturing some features from {\it linear\/}
interpolation, can be devised for Ising spins, so as to reduce
$\sigma_1$. The question is important since the Ising model, as an
extreme case, can teach us what can be done in other models far from the
Gaussian.\par
At this point the answer, as we will see, is certainly positive, although
preliminary: there are so many possibilities, and the search has just
begun. One obvious difference between constant and linear interpolation
is that the latter relates a given variable to {\it two\/} neighbors, not
one. Thus, our first attempt at a linear-like interpolation is to replace
the two-spin SW coarsening with the following {\it three spin
coarsening\/} (3SC), developed in collaboration with Dorit Ron.\par
For simplicity we describe (and have developed and tested) only the case
of uniform bonds (constant $J_{ij}$); this is not essential, but
introduces simplifying symmetries. Denote by $\beta=J_{ij}/T$ the uniform
thermal binding between neighbors. Consider a spin $s_0$ with two
neighbors, $s_-$ and $s_+$ say. The current Hamiltonian has the form
\disp{{1\over T}E=-\beta s_0s_--\beta s_0s_+-\cdots}\noindent
where the dots stand for any other terms. Three other Hamiltonians are
offered as alternatives:
\disp{\begin{array}{ll}
{1\over T}E_1=&-\infty s_0s_--as_0s_+-\cdots\\
{1\over T}E_2=&-as_0s_--\infty s_0s_+-\cdots\\
{1\over T}E_3=&-bs_-s_+-\cdots
\end{array}}\noindent
The $\infty$ value in $E_1$ $(E_2)$ means that $s_0$ and $s_-$ $(s_+)$
are blocked together. Note that in $E_3$ the two bonds between $s_0$ and
its two neighbors are deleted, but a new direct   bond is introduced
between the neighbors themselves. One selects $E_i$ with probability
$P_i$ $(i=1,2,3)$, where $P_1+P_2+P_3=1$. To obtain detailed balance,
these probabilities are taken to depend on the current value of $s_-$,
$s_0$ and $s_+$ according to Table~1 --- plus the obvious rule that $P_i(
-s_-,-s_0,-s_+)=P_i(s_-,s_0,s_+)$ --- and the value of $a$ and $b$ are
taken so that
\disp{\begin{array}{ll}
e^{2a}=&(e^{2\beta}-e^{-2\beta})/(2-2p_{\ast})\\
e^{2b}=&e^{-2\beta}/p_{\ast},\end{array}}
\begin{table}[htb]
\begin{tabular}{|ccc|c|c|c|} \hline
$s_-$&$s_0$&$s_+$&$P_1$&$P_2$&$P_3$\\ \hline
$+$&$+$&$+$&${1\over 2}(1-e^{-4\beta})$&${1\over 2}(1-e^{-4\beta})$&$e^{-
4\beta}$\\
$+$&$-$&$+$&$0$&$0$&$1$\\
$+$&$+$&$-$&$1-p_{\ast}$&$0$&$p_{\ast}$\\
$+$&$-$&$-$&$0$&$1-p_{\ast}$&$p_{\ast}$\\ \hline
\end{tabular}
\caption{}
\end{table}\noindent
$p_{\ast}$ being a small positive parameter. We chose $p_{\ast}=.15$, but
other values in the range $.05\le p_{\ast}\le .2$ are perhaps as good.
\par
The detailed balance of this, and also that of SW and other coarsening
schemes, is a special case of the following easily-proven theorem.\par
{\it Theorem\/} (Kandel-Domany \cite{Kan3}). {\it Replacing the
Hamiltonian\/ $E$ by one of the Hamiltonians\/ $E_1,\ldots,E_k$, in
probabilities\/ $P_1(s),\ldots,P_k(s)$ respectively, where\/ $s$ is the
configuration at the time of replacement, preserves detailed balance if}
\def\theequation{5.8}
\equa{P_i(s)=q_ie^{(E(s)-E_i(s))/T},}
{\it where\/ $q_i$ is independent of\/} $s$.\quad $\bbox$\par
We have tested 3SC on an $L\times L$ periodic grid by applying the
coarsening step for all triplets $s_-$, $s_0$ and $s_+$ at grid positions
$(j,2k-1)$, $(j,2k)$ and $(j,2k+1)$ respectively such that $j+k$ is even.
We compared it with an SW coarsening that terminated all the
corresponding $(s_0,s_-)$ and $(s_0,s_+)$ bonds. Results at the critical
temperature are summarized in Table~2. They show that for 3CS, unlike SW,
the ratio $\sigma_1/\chi_0$ decreases with $L$. This means that if the
susceptibility is measured on the first coarse grid, without ever
returning to the fine, the average error is small: it tends to 0 as $L$
increases.
\begin{table}[htb]
\begin{tabular}{|c|c|c|c|c|} \hline
$L$&$\chi_0$&$\sigma_0$&$\sigma_1$&$\sigma_1$\\
&&&SW&3SC\\ \hline
4&12.2&&1.8&.7\\
8&41.4&&7.2&1.5\\
16&139.5&56.8&25.6&4.0\\
32&470.2&192.5&81.6&10.6\\ \hline
\end{tabular}
\caption{}
\end{table}\par
The observation that has led to the construction of 3SC is that the basic
flaw in the SW coarsening is the introduction of many deletions, usually
clustered along well-defined lines: the lines of current boundaries of
spin alignment. These lines therefore exhibit in $E^1$ weakened
couplings, and are thus likely to persist as boundaries of spin alignment
also on coarse grids. This means strong correlation between different
coarse grid configurations. In 3SC the introduction of such
weakened-coupling lines is minimized.\par
This is just a first attempt; it all may well be done better. Observe
that the blocks created by 3SC are not necessarily {\it continguous\/}:
the Hamiltonian $E_3$ creates a bond between $s_-$ and $s_+$, so they
latter may be blocked together without having the points in between, such
as $s_0$, included in the block. More general schemes may create blocks
that are not necessarily {\it disjoint\/}. An so forth: the possibilities
are many.\par
It is not clear whether the ideal statistical efficiency is always
attainable. What {\it has\/} been established, we believe, is that it is
possible to greatly benefit from making many measurements at the coarse
levels of a multilevel Monte-Carlo algorithm, even in discrete-state
models, if a suitable coarsening scheme is used.\par
\section{Other Relevant Multilevel Techniques}
We briefly mention here several other multilevel techniques that are
relevant to lattice field computations.\par
Performing general {\it integral transforms\/}, or solving {\it integral
and integro-differential equations\/}, discretized on $n$ grid points,
have been shown to cost, using a multigrid structure, only $O(n)$ or $O(n
\log n)$ operations, even though they involve {\it full\/} $n\times n$
matrices \cite{A15}, \cite{A8}. In particular this is true for performing
Fourier transforms on {\it non-uniform\/} grids. An extension has been
devised to transforms with oscillatory kernels \cite{A8}.\par
The calculation of the $n(n-1)$ {\it interactions between\/ $n$ bodies\/}
(e.g., to obtain the residual forces in Sec.\ 1.1), can be performed in
$O(n)$ operations by embedding in a multigrid structure \cite{A8}.\par
Multilevel annealing methods have been shown as very effective for {\it
global optimization\/} of systems with a multitude of local optima and
with multi-scale attraction basins, in which cases the usual simulated
annealing method may be extremely inefficient. This includes in
particular ground-state calculations for discrete-state and frustrated
Hamiltonians \cite{A18}, \cite{A39}. Work now is in progress to extend
these multilevel annealing techniques to the calculation of ground states
of many particle problems.\par
Multilevel Monte-Carlo methods, similar to those described in Sec.~4, are
also being developed for many particle (e.g., atom) simulations. The
particles are embedded in a lattice which allow collective stochastic
moves, somewhat similar to the collective moves described in Sec.\ 1.4
above.\par
Finally, it has been demonstrated that the multigrid methodology can be
used as a tool for directly deriving the {\it macroscopic equations\/} of
a physical system. For example, in the case that the interactions (1.1)
above are Lennard-Jones or similar atomic interactions, the equations
obtained on the coarse levels of the multigrid structure described in
Sec.\ 1.6 are essentially the equations of elasticity for that material.
In this example a particle problem has given rise to a continuum
macroscopic description, expressed as partial differential equations. The
reverse can also happen: starting with PDEs, such as wave equations, the
macroscopic description may end up being that of a ray or a particle
(cf.\ Sec.\ 3.7). Sometimes, a statistical microscopic system can give
rise to a deterministic macroscopic system, or vice versa.\par
The derivation of macroscopic equations for statistical systems may be a
natural continuation of the approach described at the end of Sec.\ 5.4,
where very fine levels may be used only at very small subdomains.\par
Generally, the macroscopic equations obtained this way are expected to be
much simpler than those derived by group renormalization methods. This is
directly due to the slight ``iterativeness" left in the process by the
updates described above, which relieves the coarse level from the need to
describe in one set of equations all the possible fine-level situations.
In many cases the need for such updates may tend to disappear on
sufficiently coarse levels. Even when this is not the case, an activation
of much finer levels during large-scale (coarse level) simulations will
only rarely and locally be needed.\par
\vskip 20pt\par\noindent
{\bf Acknowledgements}\par
\bigskip
The research was supported in part by grants No.\ I-131-095.07/89 from
the German-Israeli foundation for Research and Development (GIF), No.\
399/90 from the Israel Academy of Sciences and Humanities, AFOSR-91-0156
from the United States Air Force, and NSF DMS-9015259 from the American
National Science Foundation.\par

\end{document}